\documentclass[showpacs,twocolumn,amsmath,amssymb,superscriptaddress,pra]{revtex4-1}

\usepackage{color}% color text
\usepackage{graphicx}
\usepackage{amssymb}
\usepackage{amsmath}
\usepackage{mathrsfs}
\newcommand{\ud}{\mathrm{d}}
\newcommand{\diag}{\mathrm{diag}}

\newcommand{\ket}[1]{|#1\rangle}

\DeclareMathOperator{\Tr}{Tr}

\usepackage[breaklinks=true]{hyperref}
\graphicspath{ {./figures/} {./}}

\begin{document}

\title{Vector modulational instability induced by parametric resonance in periodically tapered highly-birefringent optical fibers}

\author{Andrea Armaroli}
\email{andrea.armaroli@mpl.mpg.de}
\affiliation{Max Planck Research Group `Nonlinear Photonic Nanostructures' \\Max Planck Institute for the Science of Light, G{\"u}nther-Scharowsky-Str.~1/Bau 24
91058 Erlangen, Germany}
\author{Fabio Biancalana}
\affiliation{Max Planck Research Group `Nonlinear Photonic Nanostructures' \\Max Planck Institute for the Science of Light, G{\"u}nther-Scharowsky-Str.~1/Bau 24
91058 Erlangen, Germany}
\affiliation{School of Engineering and Physical Sciences, Heriot-Watt University, EH14 4AS Edinburgh, United Kingdom}
\date{\today}


\begin{abstract}
We study the modulational instability induced by periodic variations of group-velocity dispersion and nonlinear coefficients in a highly birefringent fiber. We observe, for each resonance order, the presence of two pairs of genuine vector type sidebands, which are spectrally unbalanced  between the polarization components for nonzero group-index mismatch, and one pair of balanced sidebands emerging and dominating at increasing  group-index mismatch. As the conventional modulational instability manifests itself, it is partially suppressed by the proximity of these new unstable regions. 
\end{abstract}

\maketitle

\section{Introduction}

In classical mechanics parametric resonance (PR) is a well-known instability phenomenon which occurs in systems the parameters of which are varied periodically during evolution \cite{ArnoldCM,LandauCM}. For example, a harmonic oscillator the frequency of which is forced to vary in time will become unstable if its internal parameters and the amplitude of the frequency variation happen to be inside special regions, known as {\em resonance tongues}. The study of the properties of resonance tongues has a long history and relies on a variety of geometrical approaches \cite{ArnoldGMODE,Broer2000}.

It is natural that such a general phenomenon was associated to the equally important instability process that is ubiquitous in infinite dimensional dynamical systems: modulation instability (MI), also known as Benjamin-Feir instability \cite{BenjaminFeir,*BespalovTalanov}. MI is known to exist in different branches of physics such as fluid-dynamics \cite{Whitham1965}, plasma physics \cite{Taniuti1968,*Hasegawa1970,*Tam1969}, Bose-Einstein condensates \cite{Konotop2001} and solid-state physics \cite{Sievers1998}. In nonlinear optics \cite{Karpman1967}, it manifests itself as pairs of sidebands  exponentially growing  on top of a plane wave initial condition, by virtue of the interplay between the cubic Kerr nonlinearity and the group velocity dispersion (GVD). In optical fibers it leads to the breakup of a plane wave into a train of normal modes of the system, i.e.~solitons \cite{Hasegawa1986,*Akhmediev1986}.

The link between PR and MI has been established  in relation to the periodic re-amplification of signals in long-haul telecommunication optical fiber cables \cite{Matera1993}. This was based on a nonlinear Schr{\"o}dinger equation (NLS) where the coefficient of the nonlinear term is varied along the propagation direction. Importantly, this peculiar type of MI occurs in both normal and anomalous GVD. This prediction was later partially verified in experiments, see \cite{Kikuchi1995}.

Moreover, in long-haul fibers, dispersion management is a commonly used technique which introduces periodic modulation of fiber characteristics. The possibility of instability phenomena disrupting adjacent communication channels has been thoroughly analyzed, see e.g.~\cite{Smith1996,Bronski1996,TchofoDinda2003,TchofoDinda2008}. Specifically, in \cite{Smith1996} the partial suppression of the conventional MI in anomalous GVD
due to a large swing dispersion management is discussed, while in \cite{Bronski1996} the degenerate case of zero average dispersion was studied. The combination of both loss and dispersion compensation is studied in \cite{TchofoDinda2003,*TchofoDinda2008}. The main interest in those works was on  step-like variations of the GVD coefficient.

At the same time the effects of smooth periodic or random variations of fiber parameters were studied in \cite{Abdullaev1996,*Abdullaev1997,Abdullaev1999}. 
Also some work has been done on the effect of the perturbation of fiber parameters on soliton propagation \cite{Bauer1995,*Pelinovsky2004}.

It turns out that the variation of dispersion and nonlinearity can enhance or suppress the PR, while higher order nonlinear effects such as self-steepening proves less important. Quite surprisingly,  experiments on  micro-structured fibers have been reported only recently for the first time, see Ref.~\cite{Droques2012,*Droques2013}, where a  photonic-crystal fiber (PCF, \cite{RussellScience2003}) of varying diameter is used. In that experiment, the dispersion is periodically switched from normal to anomalous, but this feature is not required to achieve PR, while the effect of Raman scattering plays an important role in the relative magnitude of the PR peaks. 

The conventional  explanation is in term of a grating-assisted phase matching process \cite{Matera1993, TchofoDinda2003, TchofoDinda2008, Droques2012,Droques2013}, but it was verified in Ref.~\cite{Armaroli2012} that this approximation is inaccurate if the period of parameter variation is comparable with the length scale at which the nonlinear processes occur. In Ref.~\cite{Armaroli2012} it was proved that an accurate description must be based on the Floquet theory \cite{LandauCM,ArnoldGMODE} and the use of regular perturbation techniques, such as the method of averaging \cite{VerhulstBook2010}. 

The study of birefringent fibers permits to observe a variety of new physical phenomena, which are ascribed to the presence of cross-phase modulation (XPM) terms \cite{AgrawalNL}. The MI in birefringent fibers (vector MI) occurs also for normal GVD and was extensively studied in the past not only in highly birefringent fibers (HBFs) \cite{Rothenberg1990a,Drummond1990,Rothenberg1991} but also for weak birefringence \cite{Wabnitz1988}. The effect of a step-wise variation of birefringence was considered in Ref.~\cite{Murdoch1997}, in the weakly birefringent regime, while highly-birefringent fibers with step-wise variations of dispersion were studied in Ref.~\cite{Abdullaev1999}, in a dispersion-management scenario of alternating GVD sign. This two last works apply rigorously the Floquet theory, but ignore completely the group-index mismatch. 

The possibility of tailoring the PCF birefringence, see \cite{RussellScience2003} and references therein, and of obtaining a smooth reproducible profile of fiber parameters by advanced fabrication techniques, \cite{Droques2012}, permits to achieve PR instabilities on a short distance and to explore different birefringence regimes and the effect of group velocity mismatch. 

In the present work we study parametric instabilities in a HBF with varying GVD and nonlinear coefficient. We provide accurate analytical estimate of PR peak detuning and gain and contrast them to the numerical application of Floquet theory and to split-step simulations. We observe the existence of two families of MI peaks at each PR order: one exhibits a behavior similar to conventional (i.e.~with constant parameters) vector MI {\cite{Rothenberg1990a}}, while the other resembles scalar MI and is the dominant MI process for large group-index mismatches. Finally we found that, at large group-index mismatch, the conventional vector MI is partially suppressed for large variations of parameters.

\section{Model equations and analytical estimates}

\subsection{Incoherently coupled NLS and linearized equations}

According to the conventional approach, \cite{AgrawalNL}, the propagation in HBF can be described by two incoherently-coupled NLS equations (ICNLS), which read as

\begin{multline}
	i\partial_z A_j\pm i\frac{\delta}{2} \partial_t A_j  -\frac{1}{2}\beta_2(z)\partial_{tt}A_j+\\ 
	\gamma(z)(|A_j|^2+B |A_{3-j}|^2)A_j=0\text{, with }j=1,\,2,
	%i\partial_z A_2&- i\frac{\delta}{2} \partial_t A_2  -\frac{1}{2}\beta_2(z)\partial_{tt}A_2+\gamma(z)(|A_2|^2+B |A_1|^2)A_2=0
\label{eq:CNLS1}
\end{multline}

where $\beta_2$ and $\gamma$ are normalized GVD and nonlinear coefficients, $\beta_2(z)\equiv\overline\beta_2(z)/\overline\beta_2^0$ and $\gamma(z) \equiv \overline\gamma(z)/\overline\gamma^0$; $\overline\beta_2(z)$ and $\overline\gamma(z)$ are the physical GVD and nonlinear coefficients, respectively, and the $0$ superscript denotes their mean values. $\gamma$ and $\beta_2$  are assumed to be equal for the two polarizations and periodic functions of $z$. 
Finally $z\equiv Z/Z_{\mathrm{nl}}$ is the dimensionless distance in units of the nonlinear length $Z_\mathrm{nl}\equiv(\overline\gamma^0 P_{t})^{-1}$, and $t\equiv(T-\left(v_g^0\right)^{-1} Z)/T_\mathrm{s}$ is the dimensionless retarded time in units of  $T_\mathrm{s}\equiv \sqrt{Z_{\mathrm{nl}} |\overline\beta_2^0|}$,  $v_g^0$ is the mean group velocity, and $\delta = Z_\mathrm{nl}/T_\mathrm{s}\left[(v_g^{-1})_1-(v_g^{-1})_2\right]$ is the normalized group-index mismatch between the two polarizations. $P_t$ is the total input power injected in the fiber, and $A_{1,2}$ are the dimensionless  slowly varying modal amplitudes of the two polarization components scaled by $\sqrt{P_{t}}$. 
The XPM coefficient B is used throughout the paper since the ICNLS model can be applied to other physical settings \cite{Agrawal1989,*Millot2002}. {In a HBF, the ICNLS model applies provided we set $B=2/3$, thus $A_{1,2}$ correspond to the mode polarized along the fast and slow axis, respectively.}

We look for for a steady state solution of \eqref{eq:CNLS1} in the form $A_{1,2}=\sqrt{P_{1,2}}\exp{(i \phi_{1,2}(z))}$: it can be verified that $\phi_{1,2}(z) = \left(P_{1,2}+BP_{2,1}\right)\int_{-\infty}^{z}{\gamma(z')\ud z'}$. 
We then perturb this steady state by adding a small complex time dependent contribution $a_{1,2}(z,t)$, i.e.~$A_{1,2}(z,t)=\left(\sqrt{P_{1,2}}+\varepsilon a_{1,2}(z,t)\right)\exp{(i \phi_{1,2}(z))}$, with $\varepsilon \ll 1$. Inserting this Ansatz in Eq.~\eqref{eq:CNLS1} and taking only the terms which are first order in $\varepsilon$, one finds that  $a_{1,2}$ obeys the following equation:

\begin{multline}
	i\partial_z a_{j} \pm i\frac{\delta}{2}\partial_{t}a_{j}-
	\frac{1}{2}\beta_2(z)\partial^{2}_{t}a_{j}+
	\gamma(z)\left[P_j(a_{j}+a_{j}^*)+
	\right.\\
	\left. B\sqrt{P_jP_{3-j}}(a_{3-j}+a_{3-j}^*)\right] = 0,\;j=1,\,2.
\label{eq:CNLS1lin}
\end{multline}

We further assume that the input light is polarized at an angle of $\pi/4$ with respect to the fast axis, i.e.~$P\equiv P_1=P_2=1/2$, which significantly simplifies our calculations and that GVD and nonlinearity exhibit the simplest possible periodic behavior
\begin{equation}
\begin{gathered}
	\beta_2(z) = \beta_0 + \tilde \beta(z) = \beta_0+h \beta_1\cos{\Lambda z}, \\
	\gamma(z) = \gamma_0+\tilde \gamma(z) = \gamma_0+h \gamma_1\cos{\Lambda z},
\end{gathered}
	\label{eq:dispnlcos}
\end{equation}
where generally $\beta_0=\pm1$ for normal (anomalous) GVD and $\gamma_0=1$; $\Lambda$ is the normalized spatial angular frequency for the parameter oscillations. The forcing amplitude is controlled by the parameter $h$, which must be small to guarantee the validity of our perturbative expansions. However, we find below that our estimates are reliable even for $h\sim 0.5$.
Finally we substitute in \eqref{eq:CNLS1lin} the  Ansatz
\[
a_{j}(z,t) = a_{j}^A(z) e^{-i \omega t} + a_{j}^S(z) e^{i \omega t},\;j=1,2,
\]
which permits to cast the linearized system in the  form of a 4\textsuperscript{th}-order linear ODE system
\begin{equation}
	i\frac{\ud}{\ud z}  \ket{\phi} = H(z) \ket{\phi},\;
	H(z)\equiv
	\begin{bmatrix}
		\nu &c_{1}& 0 &0\\
		{c}_{2}& \nu& 2 b& 0 \\
		0 &0 & -\nu& c_{1}\\
		2 b & 0 & {c}_{2} & -\nu 
	\end{bmatrix} 	
	\label{eq:CNLSPQ1}
\end{equation}

where 
\begin{multline}
\ket{\phi} = (u_1,v_1,u_2,v_2)^T, \\ u_j = a_j^S + a_j^{A*},\; v_j =  a_j^S - a_j^{A*},\;j=1,2
\label{eq:phasequadrature}
\end{multline}
and we defined
$
c_1(z) \equiv -\frac{\omega^2}{2}\beta_2(z) \equiv c_1^0 + h\tilde{c}_1 \cos{\Lambda z}
$,
$
c_2(z) \equiv c_{1}(z)-2\gamma(z) P \equiv c_2^0 + h\tilde{c}_2 \cos{\Lambda z}$,
$b(z) \equiv -\gamma(z) B P \equiv  b_0 + h\tilde{b} \cos{\Lambda z}$, and
$
\nu  \equiv \nu_0 \equiv -\frac{\delta}{2} \omega.
$
By replacing \eqref{eq:dispnlcos} in these definitions we can naturally split the Hamiltonian matrix $H(z)$ into average and oscillating parts, i.e. $H(z)\equiv H_0 + h\tilde{H}(z)$.

Eq.~\eqref{eq:CNLSPQ1} can be rewritten as a system of two coupled Hill's equations, i.e.~linear oscillators with periodic variation of natural frequencies, but it is more practical to deal with the original first order system directly.

\subsection{Calculating position and gain of PR peaks by the averaging method}

We first present the relation which provides the values of PR detuning. We discussed extensively in \cite{Armaroli2012} how to apply the classical theory of parametric resonance \cite{LandauCM} to problems of instability in fiber optics involving varying parameters. 

Parametric resonance is a phenomenon which is accurately described by a relation between the natural frequency of the unperturbed oscillator and the forcing term frequency. Thus we have to impose that $H_0$ has real eigenvalues, which in turn implies PR is incompatible with conventional MI, which is present in  fibers with homogeneous diameter \cite{Rothenberg1990a}. From a physical point of view this is justified by the fact that conventional MI is generally a much stronger instability effect.

%In our case this implies that Eq.~\eqref{eq:CNLSPQ1} has real eigenvalues, or equivalently PR is incompatible with conventional MI, both in a mathematical (we have damped/amplified instead of oscillating solutions) and in physical meaning, since conventional MI is generally a much stronger instability effect.

{
The choice of equal GVD and nonlinear coefficients for the two components of Eq.~\eqref{eq:CNLS1} and of the particular polarization state, see above, permits to simplify the calculation of the eigenvalues of $H_0$.
Since the matrix $H_0$ is traceless \footnote{If  $\Tr{H_0}\neq0$ we could make a change of variables to transform it into a traceless matrix}, we can write its eigenvalues as $\pm\lambda_1$ and $\pm\lambda_2$, with }
\begin{equation}
	\lambda_{1,2} = \left[c_1^0c_2^0+\nu^2 \mp 2\sqrt{\left(c_1^0\right)^2 b_0^2+c_1^0c_2^0\nu^2}\right]^{\frac{1}{2}}.
	\label{eq:EVs}
\end{equation}
A single parametric oscillator is destabilized if the unperturbed system oscillates at half an integer multiple of the forcing frequency. In the present case we have two coupled oscillators and the scenario is more complicated. We must consider four independent conditions: 
\begin{equation}
	2\lambda_{1,2} = m \Lambda,
\label{eq:Vbands}
\end{equation}
which we denote as vector MI band (V-band) and
\begin{equation}
	\lambda_1\pm\lambda_2 = m\Lambda,
\label{eq:Sbands}
\end{equation}
denoted by scalar-like MI band (S-band), where $m$ is the PR order and the reason ofor the definitions will be made clear below. 
In each case we obtain a polynomial in {the detuning $\omega_m$ of the $m$-th PR peak}. The two polynomials are reported in Appendix \ref{app:poly}.

The relations between the spatial frequency of external forcing and the eigenvalues of $H_0$ can  also be obtained by  the method of averaging \cite{VerhulstBook2010}. In its simplest formulation it is based on the method of variation of constants for inhomogeneous differential equations. This in turn is equivalent to transforming the system of Eq.~\eqref{eq:CNLSPQ1} to the interaction picture, i.e.~the evolution of the slow variables  $\ket{\phi}_I=e^{i H_{0} z}\ket{\phi}$ is governed by
\begin{equation}
	i\frac{\ud}{\ud z} \ket{\phi}_I = h  H_I(z) \ket{\phi}_I, \text{ with }
	H_I = e^{i H_{0} z} \tilde{H}(z) e^{-i H_{0} z}. 
	\label{eq:CNLSPQ2}
\end{equation}
The  averaging process is used to eliminate the remaining oscillating terms from Eq.~\eqref{eq:CNLSPQ2}. In the right-hand side we find elements with spatial periods obtained by linear combinations of $\Lambda$ and $\lambda_{1,2}$, which are in general incommensurable. Thus the method of averaging needs to be generalized by performing the integration over an infinite range, i.e.~
\begin{multline}
	i\partial_z \ket{\phi}_I = h\langle H_I(z)\rangle  \ket{\phi}_I,\;\\
	\langle H_I(z)\rangle =\lim_{Z\to\infty} \frac{1}{Z}\int_0^Z{H_I(z')\ud z'}
\end{multline}
It is clear from this matrix expression how to obtain the four PR conditions, since the above-mentioned resonances correspond to the presence of nonzero average elements in the interaction Hamiltonian. Most importantly we can estimate the peak gain, at first order in $h$, by solving for the complex  eigenvalues of the averaged interaction Hamiltonian. 

As in the scalar case of Ref.~\cite{Armaroli2012}, the  first order averaging method provides us with an estimate of the peak gain of the 1\textsuperscript{st}-order PR, which are reported in Appendix \ref{app:gain}. In order to estimate the gain of higher-order PR, a higher-order perturbation theory is demanded, but this is outside the scope of this work.

In the next paragraph we present the numerical characterization of the PR phenomenon, in the from of resonance tongues and output spectra of split-step simulations and compare it to our analytical estimates.

\section{Results and discussion}

Throughout this paragraph we set $\beta_0=+1$, normal GVD, $\Lambda=10$ and $\gamma_1=-\beta_1=1$, the latter associated to the maximum gain in the scalar case \cite{Armaroli2012}.  As a guide for our considerations we study first how the properties of PR sidebands as a function of the group-index mismatch $\delta$. This parameter was neglected in the past \cite{Abdullaev1999}, but it plays here a crucial role. For our choice of parameters, the conventional MI occurs at $\delta>\sqrt{4\beta_0\gamma_0P/3}=\sqrt{2/3}\approx0.82$.
%%%%%%%%%%%%%%%%%%%%%%%%%%%%%%%%%%%%%%%%%%%%%%%%%%%%%%%%%%%%%%%%%%%%%%%%%%%%%%%%%%%%%%%%%%%%%%%%%%%%%%%%%%%%
% Analytical estimates vs. delta 
\begin{figure}
	\centering
		\includegraphics[width=0.50\textwidth]{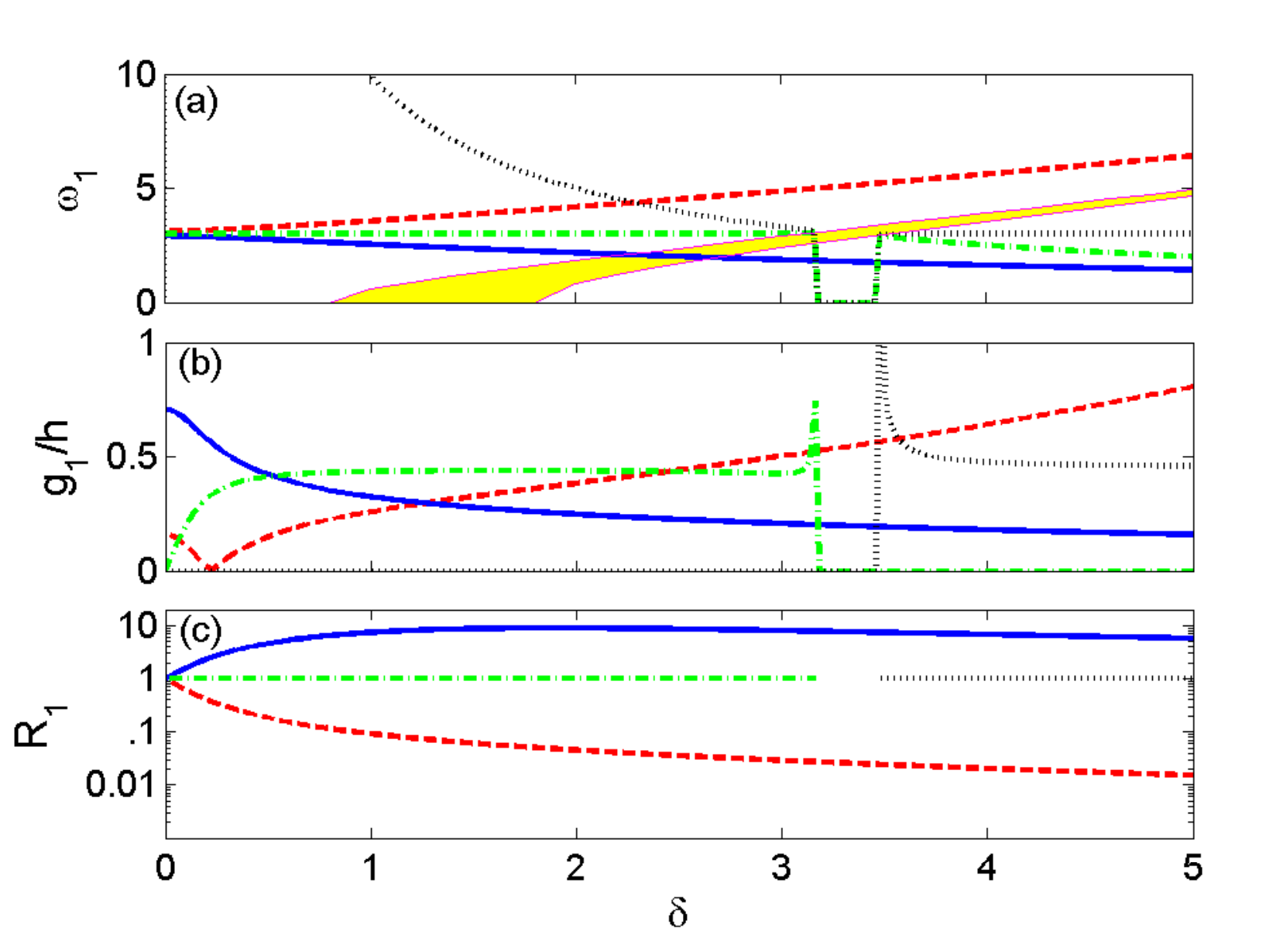}
	\caption{(Color online) Characterization of 1\textsuperscript{st} order PR as a function of normalized group-index mismatch ($\delta$), for parameters varying with $\Lambda=10$. (a) Resonant detuning; (b) Gain slope ($g_1/h$) values (c) Unbalance of V-bands around $\omega$ for the fast axis (logarithmic scale); the slow axis exhibits the opposite behavior. The S-bands are always balanced and are reported in the region of nonzero gain. The following line convention applies to every panel: blue solid line is the V-band corresponding to $\Lambda=2\lambda_2$, while the red dashed to the other V-band, $\Lambda=2\lambda_1$. The green dash-dotted line denotes the S-band $\Lambda=\lambda_2+\lambda_1$; finally the black dotted line corresponds to the other S-band, $\Lambda=\lambda_2-\lambda_1$. In panel (a), we also report the instability range of the conventional XPM MI in the absence of perturbations, as a yellow-shaded area.}
	\label{fig:figure1}
\end{figure}
%%%%%%%%%%%%%%%%%%%%%%%%%%%%%%%%%%%%%%%%%%%%%%%%%%%%%%%%%%%%%%%%%%%%%%%%%%%%%%%%%%%%%%%%%%%%%%%%%%%%%%%%%%%%%%%
In Fig.~\ref{fig:figure1} we report the analytical estimates as a function of $\delta$  of (a) the PR detuning  [Eqs.~\eqref{eq:PRcondpoly1} and \eqref{eq:PRcondpoly2}], (b) their respective gain [Eq.~\eqref{eq:gain1} and \eqref{eq:gain2}] (c) and the Stokes-antiStokes imbalance of sidebands, $R_1\equiv|a^S_1/a^{AS}_1|$, which is obtained by the eigenvectors of the averaged Hamiltonian. The imbalance is defined only for one polarization mode, as for the other polarization component is exactly the inverse, on account of the conservation of total momentum of Eq.~\eqref{eq:CNLS1}.

In Fig.~\ref{fig:figure1} we observe that the S-bands occur at constant detuning between a pair of V-bands (a), they have finite gain only if $\delta\neq0$ (b) and are spectrally symmetric around the pump frequency [see (c)], which justify our definition of scalar-like bands.  At around $\delta\approx3.5$ the conventional MI unstable sideband crosses the S-band and the latter switches from plus to minus sign in Eq.~\eqref{eq:Sbands}, as can be noticed by carefully observing the range where the gain is zero in (b). Moreover their gain is constant over a wide range of $\delta$. 

The V-bands amplitudes are perfectly symmetric around $\omega=0$ for $\delta=0$ while they develop an asymmetry for $\delta\neq0$: thus this PR bands have the same character of the conventional vector MI bands in the ICNLS system \cite{Rothenberg1990a,Biancalana2004e,Agrawal1989} and this explains our definition. Finally they are increasingly split apart as $\delta$ increases. 

The brightest, i.e.~largest gain, peak is for $\delta<0.52$ a V-band, then, for $0.52<\delta<2.5$ an S-band. We will discuss below what happens beyond $\delta\approx2.5$, where a V-band exhibits a gain larger then the S-band: the numerically computed resonance tongues shows a complicated structure where conventional MI and high-order PR coexist and the gain predictions prove inaccurate.

%We observe that V-bands  Moreover the S-bands appear only for $\delta\neq0$ as can be seen by the formula of gain but exhibits symmetric sidebands; moreover the main instability sidebands are for $\delta<0.52$ a V-band, then an S-band.

%%%%%%%%%%%%%%%%%%%%%%%%%%%%%%%%%%%%%%%%%%%%%%%%%%%%%%%%%%%%%%%%%%%%%%%%%%%%%%%%%%%%%%%%%%%%%%%%%%%%%%%%%%%%%%%%%%%%%
% Resonance tongues 1st order
\begin{figure}
	\centering
		\includegraphics[width=0.50\textwidth]{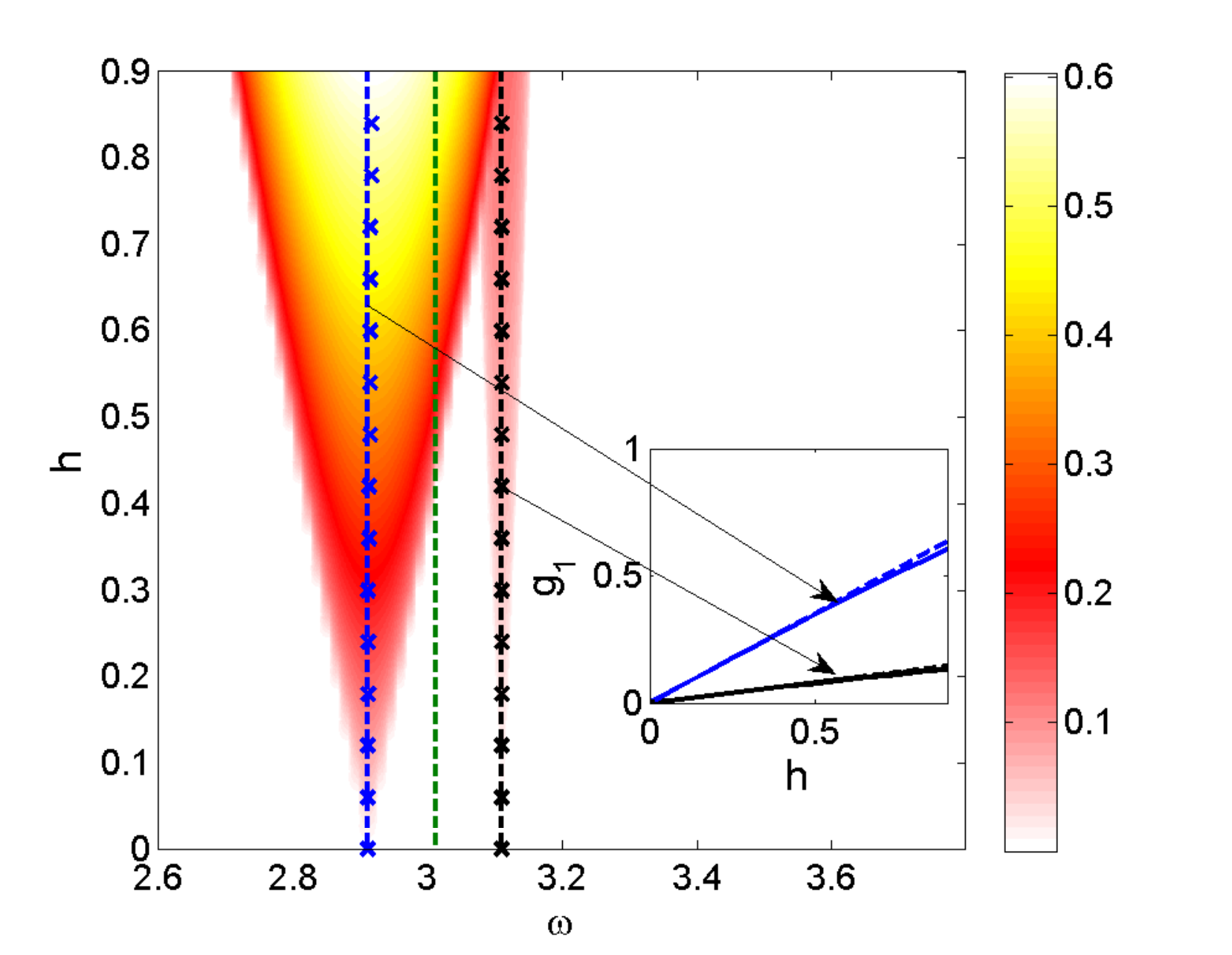}
	\caption{(Color online) Resonance tongues for 1\textsuperscript{st} order PR, with $\Lambda=10$ and $\delta=0$. The color scale corresponds to the instability gain The dashed lines denote the predicted positions of PR peaks, while the corresponding cross-marked lines represent the numerically obtained position. The corresponding maximum gain is showed in the inset (solid lines) as a function of the perturbation strength $h$ and is compared with the analytical predictions (dashed lines). The line colors in the inset correspond to those used in the $\omega\,-\,h$ diagram for the peak detuning positions.}
	\label{fig:figure2}
\end{figure}
\begin{figure}
	\centering
		\includegraphics[width=0.50\textwidth]{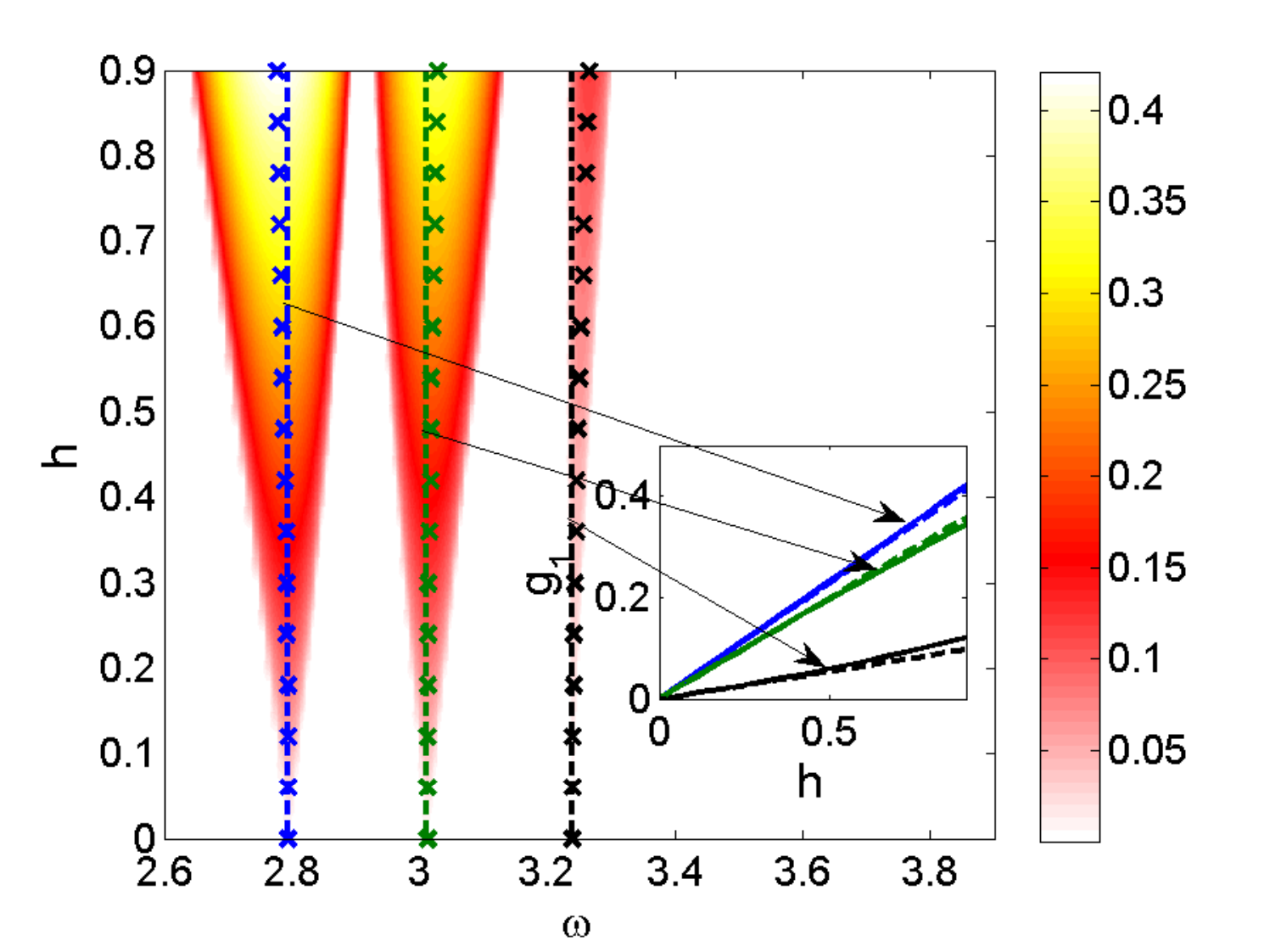}
	\caption{(Color online) Same as Fig.~\ref{fig:figure2}, with $\delta=0.4$.}
	\label{fig:figure3}
\end{figure}
\begin{figure}
	\centering
		\includegraphics[width=0.50\textwidth]{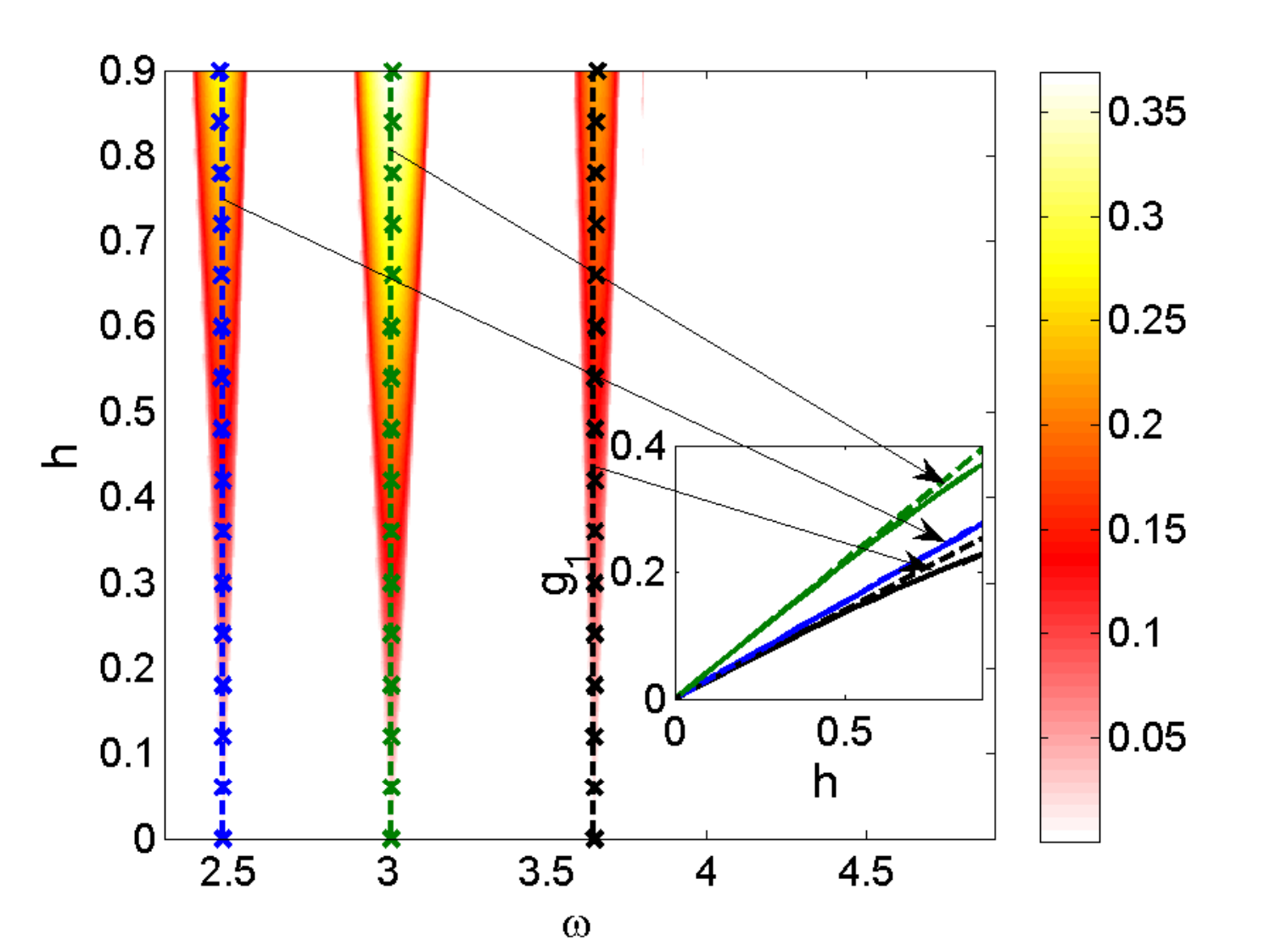}
	\caption{(Color online) Same as Fig.~\ref{fig:figure2}, with $\delta=1.15$.}
	\label{fig:figure4}
\end{figure}
\begin{figure}
	\centering
		\includegraphics[width=0.50\textwidth]{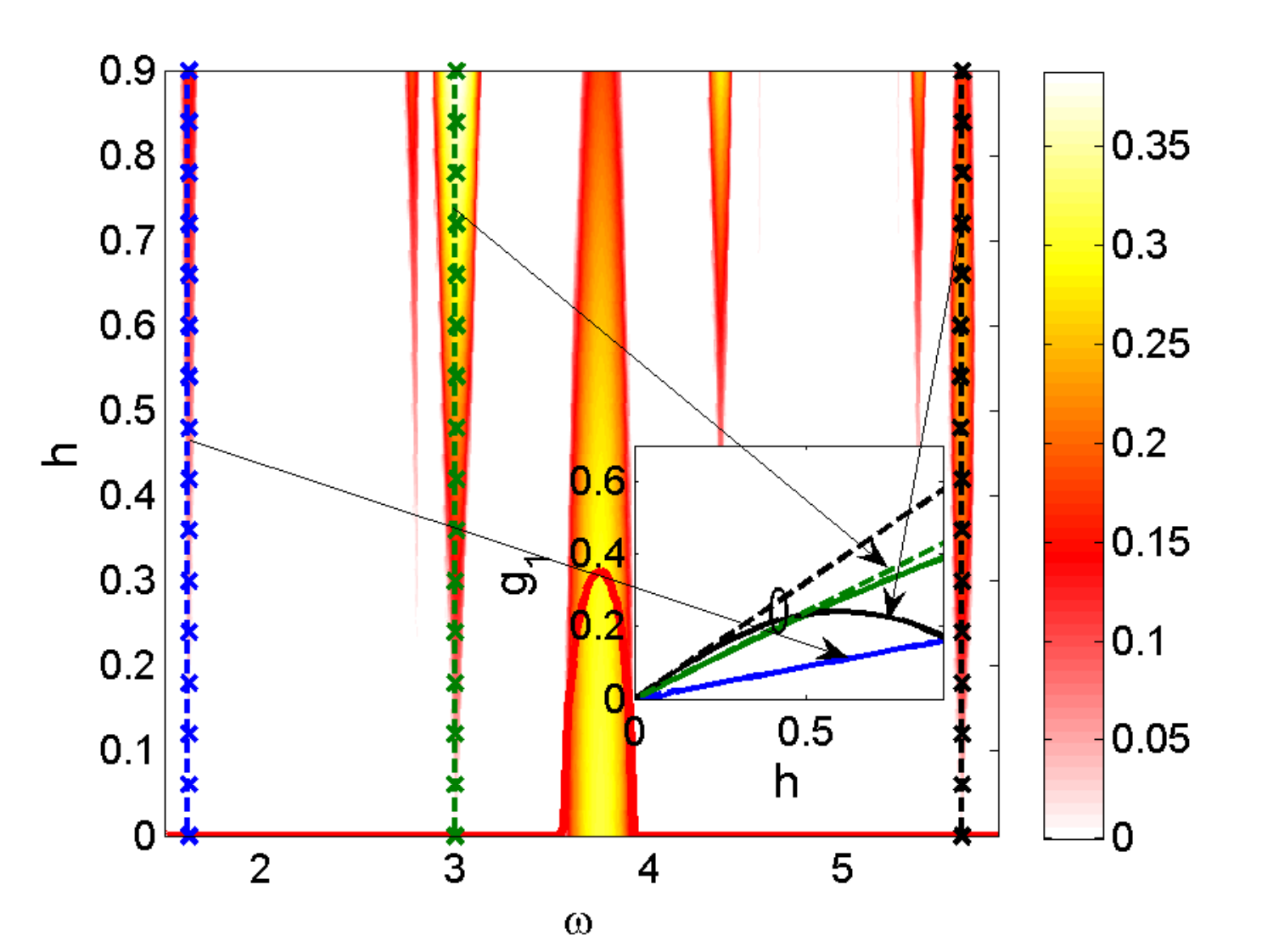}
	\caption{(Color online) Same as Fig.~\ref{fig:figure2}, with $\delta=4$. The red line shows the gain curve of conventional vector MI, which is partially suppressed at large $h$.}
	\label{fig:figure5}
\end{figure}
%%%%%%%%%%%%%%%%%%%%%%%%%%%%%%%%%%%%%%%%%%%%%%%%%%%%%%%%%%%%%%%%%%%%%%%%%%%%%%%%%%%%%%%%%%%%%%%%%%%%%%%%%%%%%%%%%%%%%%%%%%
We then discuss the structure of instability tongues { obtained by directly applying Floquet theory to Eq.~\eqref{eq:CNLSPQ1}} for 4 different cases: (i) $\delta=0$ in Fig. \ref{fig:figure2}, (ii) $\delta=0.4$  in Fig. \ref{fig:figure3} (iii) $\delta=1.15$  in Fig.~\ref{fig:figure4} and (iv) $\delta=4$  in Fig.~\ref{fig:figure5}. 

In Fig.~\ref{fig:figure2} we observe two V-bands, the first with a large peak gain and the second with a much weaker gain. Despite we consider $\delta=0$ as in \cite{Abdullaev1999}, we have the important difference that, here, the GVD varies smoothly and is always in the normal region, instead of the step-wise with alternating sign presented in that paper, so that we do not observe conventional scalar MI.  Finally we report the position of the S-band, which exhibits vanishing gain. It can be verified numerically that the V-bands grow spectrally symmetric in  both polarizations, as expected for $\delta=0$. 

Figures \ref{fig:figure3} and \ref{fig:figure4} show two similar situations: the main difference is that in the former the lower detuned V-band are the brightest MI peak while in the latter the central S-band has overcome V-bands as the brightest gain sideband. Conventional MI occurs at small $\omega$ for $\delta=1.15$, but is not reported since it is not influenced by PR and is almost independent of $h$. Before concentrating more on the case of Fig.~\ref{fig:figure3}, we finally present, in Fig.~\ref{fig:figure5}, the resonance tongues at large detuning $\delta=4$. We observe that the sideband structure of PR is still dominated by the PR S-band, while higher order sidebands [at $\omega\approx2.8$ (V), $\omega\approx4.4$ (S) of 2\textsuperscript{nd} order and  at $\omega\approx5.4$ (S) of  3\textsuperscript{rd} order] are interleaved with the 1\textsuperscript{st} order ones. Moreover the conventional MI and the V-band at $\omega=5.6$ are partially suppressed for large $h$, on account of the proximity of the higher-order peaks. At such values the first order estimate of gain is clearly inadequate (see inset), as in general occurs for the V-bands which coexist at large detuning with the conventional MI and higher-order PR peaks. We thus observe that for $\delta>3.5$, where the conventional MI  occurs beyond the brightest PR peaks of scalar-like type,  the variations of parameters enhances spectrally symmetric, scalar-like, sidebands and suppresses the  asymmetric sidebands which are commonly considered as the characterizing feature of MI in HBFs.

{ This behavior is consistent with the suppression of vector MI sidebands due to fluctuations of the fiber parameters, which has for long precluded their observation in PCFs, see \cite{Kibler2004,*Wong2005,*Chen2006,*Kudlinski2013}. 
We tested the effect of periodic variations of $\delta$ and observe a reduction of the peak gain of V-bands for large average $\delta$. Expressions of gain can be obtained, but are more involved than those presented in Appendix \ref{app:gain}; the PR detuning values are robust to this perturbations and the instability growth happens on a length scale larger than the period of the parameter variations, so we decided not to explicitly consider variations of $\delta$ here.
}

%%%%%%%%%%%%%%%%%%%%%%%%%%%%%%%%%%%%%%%%%%%%%%%%%%%%%%%%%%%%%%%%%%%%%%%%%%%%%%%%%%%%%
% Example of 2nd order PR
% Tongues
\begin{figure}
	\centering
		\includegraphics[width=0.50\textwidth]{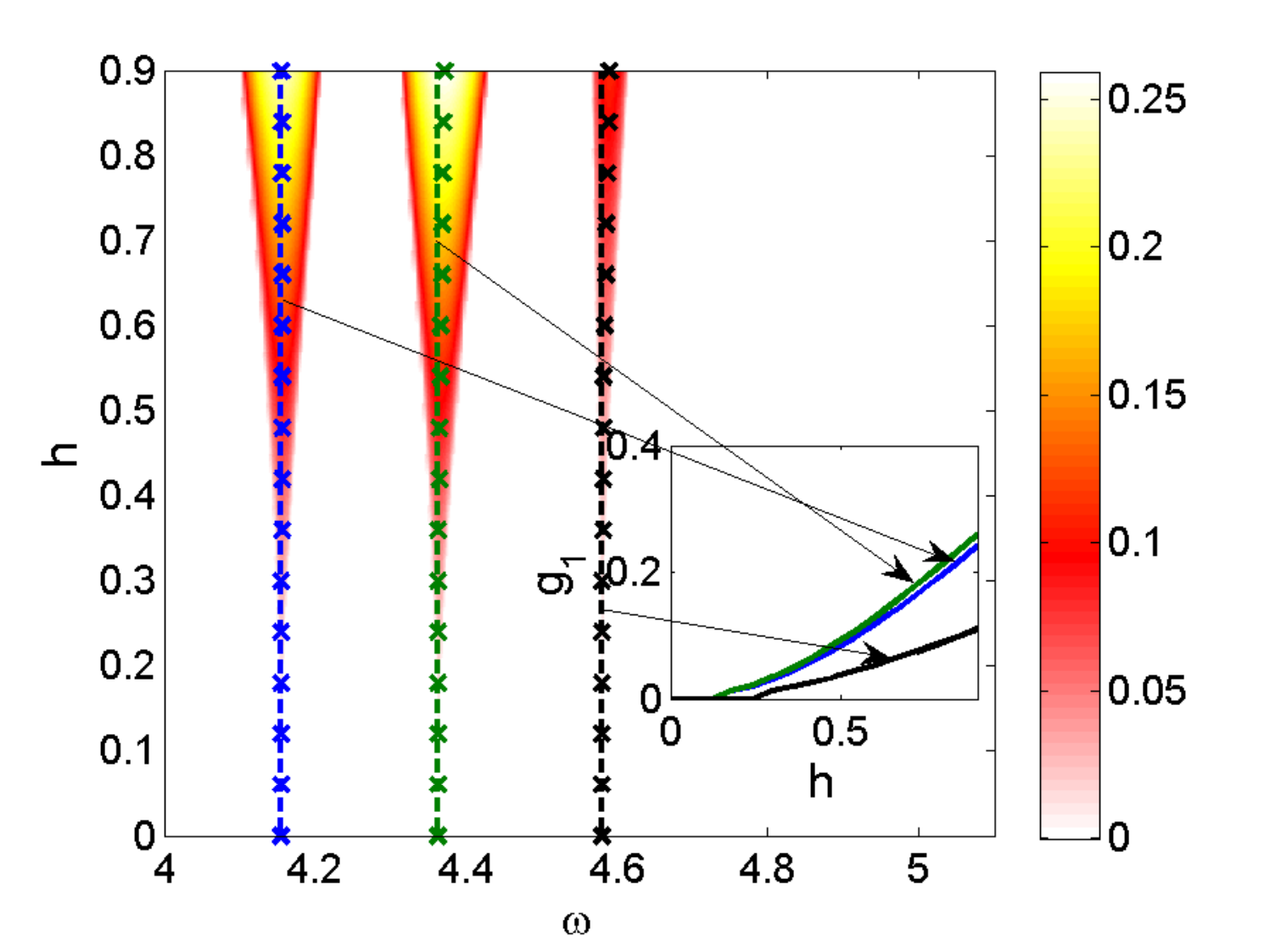}
	\caption{(Color online) 2\textsuperscript{nd}-order PR instability regions, same parameters and conventions as in Fig.~\ref{fig:figure3}. In the inset only the numerical results are reported, since  analytical estimates are not considered here.}
	\label{fig:figure6}
\end{figure}
We now complete the characterization in the case of Fig.~\ref{fig:figure3}, by showing in Fig.~\ref{fig:figure6}  the second order PR instability regions for $\delta=0.4$, which share  the same features of the 1\textsuperscript{st} order ones, except the central S-band gain is already slightly larger than the smaller detuned V-band. 

%%%%%%%%%%%%%%%%%%%%%%%%%%%%%%%%%%%%%%%%%%%%%%%%%%%%%%%%%%%%%%%%%%%%%%%%%%%%%%%%%%%%%%%%%%%%%%%%%%%%%%%%%%%%%%%%%%
% Split step
\begin{figure}
	\centering
		\includegraphics[width=0.50\textwidth]{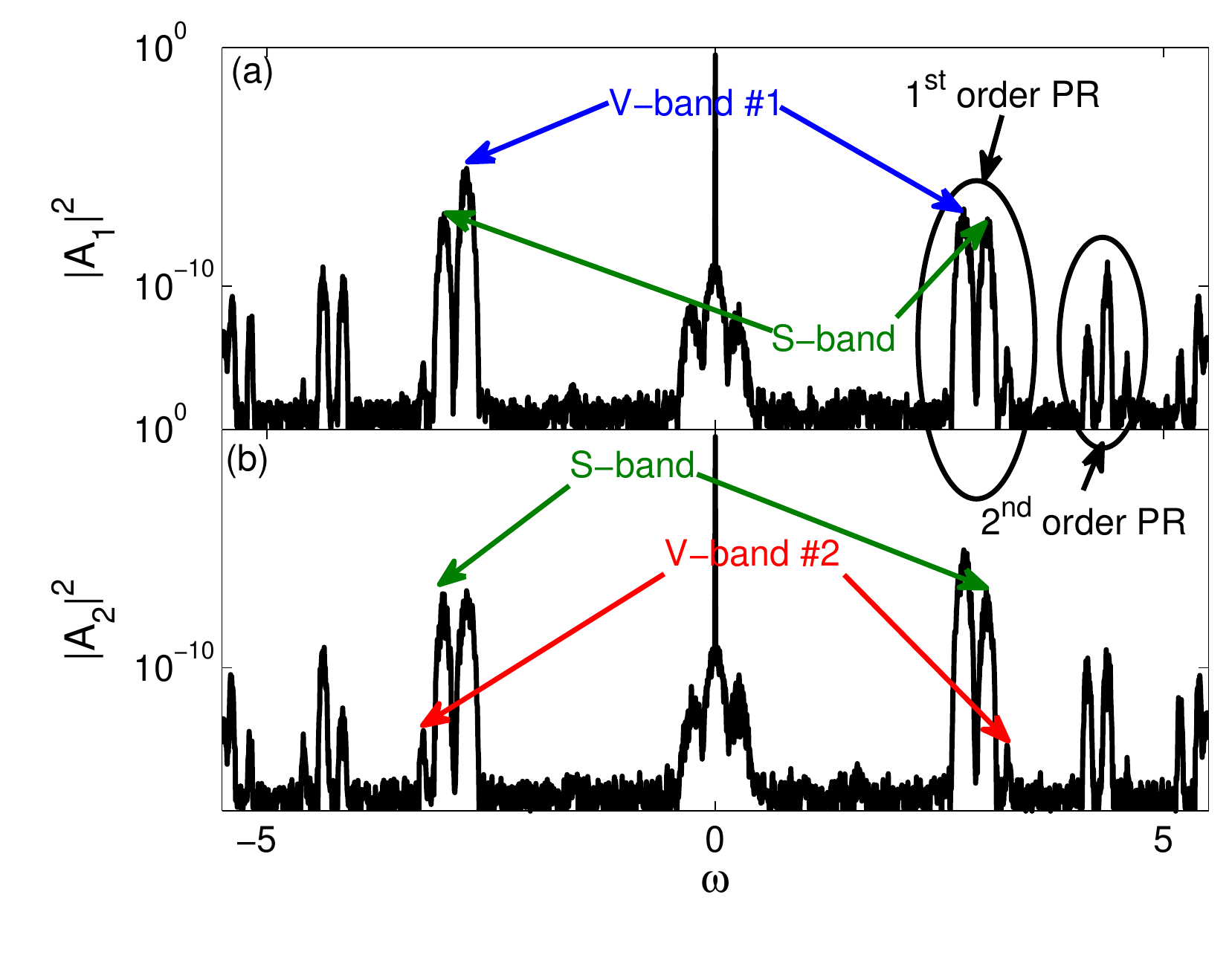}
	\caption{(Color online) Output intensity spectrum (in logarithmic scale) at normalized distance $z=30$ obtained by split-step numerical simulation. $h=0.9$, $\delta=0.4$ and the other parameters are as in the previous figures; (a) axis 1 and (b) axis 2. We identify and classify the 1\textsuperscript{st} and 2\textsuperscript{nd} order PR peaks, as indicated by the text in the panel, by their spectral imbalance with respect to $\omega=0$. The small peaks near the pump components are the four-wave mixing product of the brightest 1\textsuperscript{st}-order V-band and S-band. }
	\label{fig:figure7}
\end{figure}
\begin{figure}
	\centering
		\includegraphics[width=0.50\textwidth]{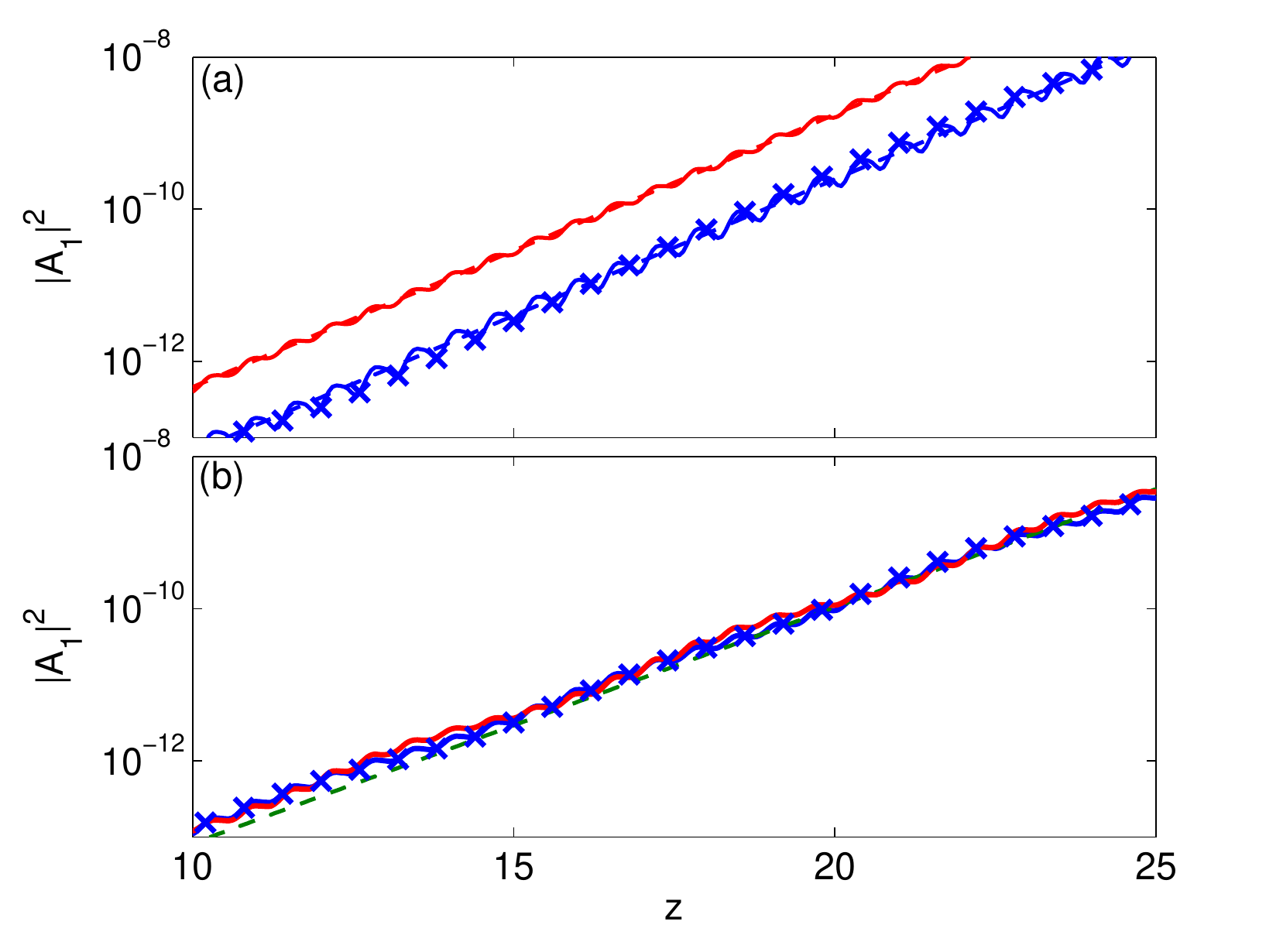}
	\caption{(Color online) Simulated evolution, on the fast axis, of the intensity of the two main peak frequencies of 1\textsuperscript{st} order PR:  (a) V-band at $\omega=2.79$  and (b) the S-band $\omega=3.01$. Red (blue with crosses) solid lines corresponds to the numerical  evolution of  (anti-)Stokes-sideband; dashed line are the average growth predicted by analytical calculations.}
	\label{fig:figure8}
\end{figure}
%%%%%%%%%%%%%%%%%%%%%%%%%%%%%%%%%%%%%%%%%%%%%%%%%%%%%%%%%%%%%%%%%%%%%%%%%%%%%%%%%%%%%
Finally we include the  output spectra obtained by solving the system of Eq.~\eqref{eq:CNLS1} by means of the split-step method, see Fig.~\ref{fig:figure7}. We set all the parameters as above, $h=0.9$ and $\delta=0.4$. 

We clearly identify the 1\textsuperscript{st} and 2\textsuperscript{nd} order PR, each of which is composed by three peaks. It is thus clear that the S-bands occur between a couple of V-bands and are symmetrically growing about the pump in both axis. The two V-bands exhibit an imbalance about $\omega=0$ which is reversed from fast to slow axis on account of the conservation of total momentum for the model of Eq.~\eqref{eq:CNLS1}, compare panels (a) and (b). Moreover the two V-bands exhibit opposite symmetry; consider the fast axis Fig.~\ref{fig:figure7}(a): the first peak at each order is characterized by the Stokes sideband outgrowing the anti-Stokes, while the second exhibits the opposite behavior---the anti-Stokes dominates over the Stokes. This is analogous to what occurs in the proximity of zero dispersion due to the presence of higher-order dispersion \cite{Biancalana2004e}.

The growth trend is presented in Fig.~\ref{fig:figure8}, for the two brightest first order peaks. The exponential growth of the unstable frequencies is superimposed to an oscillation at spatial (angular) frequency $\Lambda$, as in the scalar PR \cite{Armaroli2012}; this is quite effectively explained by the theory of averaging. However there is a remarkable difference between the two peaks: (a) the V-band involves only one eigenvalue ($\lambda_2=\Lambda/2$) of $H_0$ and grows upon a simple oscillation, while (b) the S-band involves both eigenvalues of $H_0$ and exhibits a beat of the fast  oscillations, corresponding to $\lambda_1+\lambda_2=\Lambda$, and slow oscillations, corresponding to $\lambda_2-\lambda_1=\Lambda$. 

Finally the imbalance as computed numerically from Fig.~\ref{fig:figure8}(a) is smaller than that in Fig.~\ref{fig:figure1}(c);  this a general trend: we observed that for large $h$ the imbalance of V-bands is smaller than expected by the eigenvectors of the averaged interaction Hamiltonian. 

\section{Conclusions}

In this paper we studied the effect of  the periodic variation of group-velocity dispersion and nonlinearity on  the propagation of light in a highly-birefringent optical fiber. We showed that MI sidebands  are effectively described in terms of parametric instabilities of a system of coupled oscillators and provide accurate analytical estimates of their detuning and gain. We considered only the normal GVD regime and discovered the existence of two different kinds of unstable sidebands: the first, similar to conventional MI, appears as two pairs of sidebands which generally exhibit spectral imbalance around the pumps, while the second manifests itself only for non-zero group-index mismatch as a pair of spectrally balanced peaks: moreover the latter becomes the brightest unstable peak for large enough mismatch values. The vector sidebands and the conventional vector (XPM) MI  are partially suppressed by the proximity of scalar-like  PR sidebands. The PR peak position is widely tunable by varying the period of variation of parameters and the input power; this phenomenon could thus find interesting applications in quantum optics. 

\section*{Acknowledgments}

The authors acknowledge fruitful discussions with Arnaud Mussot and Alexandre Kudlinski, Universit{\'e} Lille 1, Laboratoire PhLAM, and the financial support of the German Max Planck Society for the advancement of Science.

\appendix

\section{Estimate of resonant detuning: expression of polynomials}
\label{app:poly}
In order to obtain the resonant detuning we substitute all the quantities defined after Eq.~\eqref{eq:phasequadrature}  in  Eq.~\eqref{eq:EVs}, then recast the conditions for $m$-th order PR as a polynomial in $\omega_m^2$, which read as 
\begin{equation}
\begin{aligned}
P_1(\omega^2)=&\frac{\beta_0^4}{16} \omega^8+\frac{1}{16} (8 P \beta_0^3 \gamma_0 - 
    2 \beta_0^2 \delta^2) \omega^6 + \\
    &\frac{1}{16} \left[-2 m^2 \Lambda^2 \beta_0^2 - 
    16 (-1 + B^2) P^2 \beta_0^2 \gamma_0^2 - \right.\\
    &\left.8 P \beta_0 \gamma_0 \delta^2 + \delta^4\right] \omega^4+\\
		&\frac{1}{16} \left[-8 m^2  \Lambda^2 \beta_0 \gamma_0P - 
    2 m^2 \Lambda^2 \delta^2\right] \omega^2+\\
    &\frac{m^4 \Lambda^4}{16}, 
\label{eq:PRcondpoly1}
\end{aligned}
\end{equation}
for V-bands, Eq.~\eqref{eq:Vbands}, and
\begin{equation}
\begin{aligned}
P_2(\omega^2) =& \frac{1}{4} \beta_0 ^2 \delta ^2 \omega ^6+\\
	&\frac{1}{4}  \left[4 \beta_0 ^2 B^2 \gamma_0 ^2 P^2-\beta_0 ^2 \Lambda^2 m^2+4 \beta_0  \gamma_0 P \delta ^2 \right]\omega ^4+\\
	&\frac{1}{4} \left[\delta ^2 \Lambda ^2 \left(-m^2\right)-4 \beta_0  \gamma_0 P m^2 \Lambda ^2  \right] \omega ^2+\frac{\Lambda ^4 m^4}{4}, 
\label{eq:PRcondpoly2}
\end{aligned}
\end{equation}
for S-bands, Eq.~\eqref{eq:Sbands}.

\section{Method of averaging: instability gain}
\label{app:gain}

The explicit calculation leading expressions of the PR peak gain is quite tedious, thus we summarize here the main points and results. Let us diagonalize $H_0$ as
\[
	H_0 = V\Delta V^{-1}
\]
where $V$ is the matrix, the columns of which are the eigenvectors of $H_0$ (not necessarily normalized) and $\Delta=\diag\left[-\lambda_2,-\lambda_1,\lambda_1,\lambda_2\right]$.

The interaction Hamiltonian in Eq.~\eqref{eq:CNLSPQ2} can be expanded as
\[
H_I = V \exp(i \Delta z) V^{-1} \tilde{H} V \exp(-i \Delta z) V^{-1}
\]
but, since $V$ does not depend on $z$, we can resort to the similar matrix
\[
H_I' = \exp(i \Delta z) V^{-1} \tilde{H} V \exp(-i \Delta z),
\]
thus simplifying the resulting averaged matrix.

The expressions of gain are different for each of the cases in Eq.~\eqref{eq:Vbands} and \eqref{eq:Sbands}; 
we express them in compact form as
\begin{widetext}
\begin{equation}
%\begin{aligned}
	g_1^V = \left|\frac{h}{4 c_1^0 b_0 \lambda _2 \left(\lambda _2^2-\lambda _1^2\right)}\right|
	\left\{
	\left[A_1\tilde{c}_1+B_1\tilde{c}_2+C_1\tilde{b}\right]
	\left[A_2\tilde{c}_1+B_2\tilde{c}_2+C_2\tilde{b}\right]
	\right\}^\frac{1}{2}	
\label{eq:gain1}
%\end{aligned}
\end{equation}
for Eq.~\eqref{eq:Vbands}, with $2\lambda_2=\Lambda$, with $A_1 = b_0\left[4c_1^0c_2^0\nu(\nu+\lambda_2)+(\lambda_2^2-\lambda_1^2)(\lambda_2+\nu)^2\right]$, 
$A_2 = b_0\left[4c_1^0c_2^0\nu(\nu-\lambda_2)+(\lambda_2^2-\lambda_1^2)(\lambda_2-\nu)^2\right]$,
$B_1 = b_0c_1^0\left[\left(\lambda_2+2\nu\right)^2-\lambda_1^2\right]$,
$B_2 = b_0c_1^0\left[\left(\lambda_2-2\nu\right)^2-\lambda_1^2\right]$,
$C_1 = 4b_0^2\left(c_1^0\right)^3-c_1^0\left[c_1^0c_2^0-\left(\lambda_2+\nu\right)^2\right]
\left(c_1^0c_2^0-\lambda_1^2+3\nu^2+2\lambda_2\nu\right)$,
$C_2 = 4b_0^2\left(c_1^0\right)^3-c_1^0\left[c_1^0c_2^0-\left(\lambda_2-\nu\right)^2\right]
\left(c_1^0c_2^0-\lambda_1^2+3\nu^2-2\lambda_2\nu\right)$. 
\end{widetext}

In order to obtain the gain of the other V-band, $2\lambda_1=\Lambda$, we must replace $\lambda_2\mapsto\lambda_1$.

The peak gain of the S-band of Eq.~\eqref{eq:Sbands}, for $\lambda_1+\lambda_2=\Lambda$, is expressed by
\begin{widetext}
\begin{equation}
%\begin{aligned}
	g_1^S = \left|\frac{h}{4  b_0 \lambda _2 \left(\lambda _2^2-\lambda _1^2\right)}\right|\left[\frac{1}{\lambda_1\lambda_2}\right]^\frac{1}{2}	
	\left\{
	\left[D_1\left(\tilde{c}_1c_2^0-\tilde{c}_2c_1^0\right)+E_1\tilde{b}\right]
	\left[D_2\left(\tilde{c}_1c_2^0-\tilde{c}_2c_1^0\right)+E_2\tilde{b}\right]
	\right\}^\frac{1}{2}	
\label{eq:gain2}
%\end{aligned}
\end{equation}
with
$D_1=2\nu (2\nu-\Lambda)$, $D_2=2\nu (2\nu+\Lambda)$, 
$E_1 =\left(c_1^0c_2^0-(\lambda_1-\nu)^2\right)\left(c_1^0c_2^0-\lambda_1^2+3\nu^2-2\lambda_2\nu\right)-4b_0^2\left(c_1^0\right)^2$,
$E_2 =\left(c_1^0c_2^0-(\lambda_2+\nu)^2\right)\left(c_1^0c_2^0-\lambda_2^2+3\nu^2+2\lambda_1\nu\right)-4b_0^2\left(c_1^0\right)^2$.

Instead, 
for $\lambda_2-\lambda_1=\Lambda$,  the expression in Eq.~\eqref{eq:gain2} is valid provided the new coefficients, 
denoted by a prime, are used instead of the unprimed ones,
$D_1'=-2\nu (2\nu-\Lambda)$, $D_2'=-2\nu (2\nu+\Lambda)$, 
$E_1' =-\left(c_1^0c_2^0-(\lambda_1+\nu)^2\right)\left(c_1^0c_2^0-\lambda_1^2+3\nu^2-2\lambda_2\nu\right)+4b_0^2\left(c_1^0\right)^2$,
$E_2' =\left(c_1^0c_2^0-(\lambda_2+\nu)^2\right)\left(c_1^0c_2^0-\lambda_2^2+3\nu^2-2\lambda_1\nu\right)+4b_0^2\left(c_1^0\right)^2$.
\end{widetext}

\bibliography{ParametricResonance0313}

\begin{thebibliography}{45}%
\makeatletter
\providecommand \@ifxundefined [1]{%
 \@ifx{#1\undefined}
}%
\providecommand \@ifnum [1]{%
 \ifnum #1\expandafter \@firstoftwo
 \else \expandafter \@secondoftwo
 \fi
}%
\providecommand \@ifx [1]{%
 \ifx #1\expandafter \@firstoftwo
 \else \expandafter \@secondoftwo
 \fi
}%
\providecommand \natexlab [1]{#1}%
\providecommand \enquote  [1]{``#1''}%
\providecommand \bibnamefont  [1]{#1}%
\providecommand \bibfnamefont [1]{#1}%
\providecommand \citenamefont [1]{#1}%
\providecommand \href@noop [0]{\@secondoftwo}%
\providecommand \href [0]{\begingroup \@sanitize@url \@href}%
\providecommand \@href[1]{\@@startlink{#1}\@@href}%
\providecommand \@@href[1]{\endgroup#1\@@endlink}%
\providecommand \@sanitize@url [0]{\catcode `\\12\catcode `\$12\catcode
  `\&12\catcode `\#12\catcode `\^12\catcode `\_12\catcode `\%12\relax}%
\providecommand \@@startlink[1]{}%
\providecommand \@@endlink[0]{}%
\providecommand \url  [0]{\begingroup\@sanitize@url \@url }%
\providecommand \@url [1]{\endgroup\@href {#1}{\urlprefix }}%
\providecommand \urlprefix  [0]{URL }%
\providecommand \Eprint [0]{\href }%
\providecommand \doibase [0]{http://dx.doi.org/}%
\providecommand \selectlanguage [0]{\@gobble}%
\providecommand \bibinfo  [0]{\@secondoftwo}%
\providecommand \bibfield  [0]{\@secondoftwo}%
\providecommand \translation [1]{[#1]}%
\providecommand \BibitemOpen [0]{}%
\providecommand \bibitemStop [0]{}%
\providecommand \bibitemNoStop [0]{.\EOS\space}%
\providecommand \EOS [0]{\spacefactor3000\relax}%
\providecommand \BibitemShut  [1]{\csname bibitem#1\endcsname}%
\let\auto@bib@innerbib\@empty
%</preamble>
\bibitem [{\citenamefont {Arnold}\ \emph {et~al.}(1989)\citenamefont {Arnold},
  \citenamefont {Weinstein},\ and\ \citenamefont {Vogtmann}}]{ArnoldCM}%
  \BibitemOpen
  \bibfield  {author} {\bibinfo {author} {\bibfnamefont {V.~I.}\ \bibnamefont
  {Arnold}}, \bibinfo {author} {\bibfnamefont {A.}~\bibnamefont {Weinstein}}, \
  and\ \bibinfo {author} {\bibfnamefont {K.}~\bibnamefont {Vogtmann}},\ }\href
  {http://www.amazon.com/Mathematical-Classical-Mechanics-Graduate-Mathematics/dp/0387968903%3FSubscriptionId%3D0JYN1NVW651KCA56C102%26tag%3Dtechkie-20%26linkCode%3Dxm2%26camp%3D2025%26creative%3D165953%26creativeASIN%3D0387968903}
  {\emph {\bibinfo {title} {Mathematical Methods of Classical Mechanics
  (Graduate Texts in Mathematics)}}}\ (\bibinfo  {publisher} {Springer},\
  \bibinfo {year} {1989})\BibitemShut {NoStop}%
\bibitem [{\citenamefont {Landau}\ and\ \citenamefont
  {Lifshitz}(1976)}]{LandauCM}%
  \BibitemOpen
  \bibfield  {author} {\bibinfo {author} {\bibfnamefont {L.~D.}\ \bibnamefont
  {Landau}}\ and\ \bibinfo {author} {\bibfnamefont {E.}~\bibnamefont
  {Lifshitz}},\ }\href
  {http://www.amazon.com/Mechanics-Third-Edition-Theoretical-Physics/dp/0750628960%3FSubscriptionId%3D0JYN1NVW651KCA56C102%26tag%3Dtechkie-20%26linkCode%3Dxm2%26camp%3D2025%26creative%3D165953%26creativeASIN%3D0750628960}
  {\emph {\bibinfo {title} {Mechanics, Third Edition: Volume 1 (Course of
  Theoretical Physics)}}}\ (\bibinfo  {publisher} {Butterworth-Heinemann},\
  \bibinfo {year} {1976})\BibitemShut {NoStop}%
\bibitem [{\citenamefont {Arnold}(1988)}]{ArnoldGMODE}%
  \BibitemOpen
  \bibfield  {author} {\bibinfo {author} {\bibfnamefont {V.}~\bibnamefont
  {Arnold}},\ }\href
  {http://www.amazon.com/Geometrical-Differential-Grundlehren-mathematischen-Wissenschaften/dp/0387966498%3FSubscriptionId%3D0JYN1NVW651KCA56C102%26tag%3Dtechkie-20%26linkCode%3Dxm2%26camp%3D2025%26creative%3D165953%26creativeASIN%3D0387966498}
  {\emph {\bibinfo {title} {Geometrical Methods in the Theory of Ordinary
  Differential Equations (Grundlehren der mathematischen Wissenschaften) (v.
  250)}}}\ (\bibinfo  {publisher} {Springer},\ \bibinfo {year}
  {1988})\BibitemShut {NoStop}%
\bibitem [{\citenamefont {Broer}\ and\ \citenamefont
  {Sim{\'o}}(2000)}]{Broer2000}%
  \BibitemOpen
  \bibfield  {author} {\bibinfo {author} {\bibfnamefont {H.~W.}\ \bibnamefont
  {Broer}}\ and\ \bibinfo {author} {\bibfnamefont {C.}~\bibnamefont
  {Sim{\'o}}},\ }\href {\doibase 10.1006/jdeq.2000.3804} {\bibfield  {journal}
  {\bibinfo  {journal} {J. Differ. Equations}\ }\textbf {\bibinfo {volume}
  {166}},\ \bibinfo {pages} {290} (\bibinfo {year} {2000})}\BibitemShut
  {NoStop}%
\bibitem [{\citenamefont {{Benjamin}}\ and\ \citenamefont
  {{Feir}}(1967)}]{BenjaminFeir}%
  \BibitemOpen
  \bibfield  {author} {\bibinfo {author} {\bibfnamefont {T.~B.}\ \bibnamefont
  {{Benjamin}}}\ and\ \bibinfo {author} {\bibfnamefont {J.~E.}\ \bibnamefont
  {{Feir}}},\ }\href {\doibase 10.1017/S002211206700045X} {\bibfield  {journal}
  {\bibinfo  {journal} {J. Fluid Mech.}\ }\textbf {\bibinfo {volume} {27}},\
  \bibinfo {pages} {417} (\bibinfo {year} {1967})}\BibitemShut {NoStop}%
\bibitem [{\citenamefont {{Bespalov}}\ and\ \citenamefont
  {{Talanov}}(1966)}]{BespalovTalanov}%
  \BibitemOpen
  \bibfield  {author} {\bibinfo {author} {\bibfnamefont {V.~I.}\ \bibnamefont
  {{Bespalov}}}\ and\ \bibinfo {author} {\bibfnamefont {V.~I.}\ \bibnamefont
  {{Talanov}}},\ }\href@noop {} {\bibfield  {journal} {\bibinfo  {journal}
  {Pis'ma Zh.~Eksp.~Teor. Fiz. [JETP Letters]}\ }\textbf {\bibinfo {volume}
  {3}},\ \bibinfo {pages} {307} (\bibinfo {year} {1966})}\BibitemShut {NoStop}%
\bibitem [{\citenamefont {Whitham}(1965)}]{Whitham1965}%
  \BibitemOpen
  \bibfield  {author} {\bibinfo {author} {\bibfnamefont {G.~B.}\ \bibnamefont
  {Whitham}},\ }\href {\doibase 10.1098/rspa.1965.0019} {\bibfield  {journal}
  {\bibinfo  {journal} {Proc. Phys. Soc. London, Sect. A}\ }\textbf {\bibinfo
  {volume} {283}},\ \bibinfo {pages} {238} (\bibinfo {year}
  {1965})}\BibitemShut {NoStop}%
\bibitem [{\citenamefont {Taniuti}\ and\ \citenamefont
  {Washimi}(1968)}]{Taniuti1968}%
  \BibitemOpen
  \bibfield  {author} {\bibinfo {author} {\bibfnamefont {T.}~\bibnamefont
  {Taniuti}}\ and\ \bibinfo {author} {\bibfnamefont {H.}~\bibnamefont
  {Washimi}},\ }\href {\doibase 10.1103/PhysRevLett.21.209} {\bibfield
  {journal} {\bibinfo  {journal} {\prl}\ }\textbf {\bibinfo {volume} {21}},\
  \bibinfo {pages} {209} (\bibinfo {year} {1968})}\BibitemShut {NoStop}%
\bibitem [{\citenamefont {Hasegawa}(1970)}]{Hasegawa1970}%
  \BibitemOpen
  \bibfield  {author} {\bibinfo {author} {\bibfnamefont {A.}~\bibnamefont
  {Hasegawa}},\ }\href {\doibase 10.1103/PhysRevLett.24.1165} {\bibfield
  {journal} {\bibinfo  {journal} {\prl}\ }\textbf {\bibinfo {volume} {24}},\
  \bibinfo {pages} {1165} (\bibinfo {year} {1970})}\BibitemShut {NoStop}%
\bibitem [{\citenamefont {Tam}(1969)}]{Tam1969}%
  \BibitemOpen
  \bibfield  {author} {\bibinfo {author} {\bibfnamefont {C.~K.~W.}\
  \bibnamefont {Tam}},\ }\href {\doibase 10.1063/1.2163663} {\bibfield
  {journal} {\bibinfo  {journal} {Phys. Fluids}\ }\textbf {\bibinfo {volume}
  {12}},\ \bibinfo {pages} {1028} (\bibinfo {year} {1969})}\BibitemShut
  {NoStop}%
\bibitem [{\citenamefont {Abdullaev}\ \emph {et~al.}(2001)\citenamefont
  {Abdullaev}, \citenamefont {Baizakov}, \citenamefont {Darmanyan},
  \citenamefont {Konotop},\ and\ \citenamefont {Salerno}}]{Konotop2001}%
  \BibitemOpen
  \bibfield  {author} {\bibinfo {author} {\bibfnamefont {F.~K.}\ \bibnamefont
  {Abdullaev}}, \bibinfo {author} {\bibfnamefont {B.~B.}\ \bibnamefont
  {Baizakov}}, \bibinfo {author} {\bibfnamefont {S.~A.}\ \bibnamefont
  {Darmanyan}}, \bibinfo {author} {\bibfnamefont {V.~V.}\ \bibnamefont
  {Konotop}}, \ and\ \bibinfo {author} {\bibfnamefont {M.}~\bibnamefont
  {Salerno}},\ }\href {\doibase 10.1103/PhysRevA.64.043606} {\bibfield
  {journal} {\bibinfo  {journal} {\pra}\ }\textbf {\bibinfo {volume} {64}},\
  \bibinfo {pages} {043606} (\bibinfo {year} {2001})}\BibitemShut {NoStop}%
\bibitem [{\citenamefont {Lai}\ and\ \citenamefont
  {Sievers}(1998)}]{Sievers1998}%
  \BibitemOpen
  \bibfield  {author} {\bibinfo {author} {\bibfnamefont {R.}~\bibnamefont
  {Lai}}\ and\ \bibinfo {author} {\bibfnamefont {A.~J.}\ \bibnamefont
  {Sievers}},\ }\href {\doibase 10.1103/PhysRevB.57.3433} {\bibfield  {journal}
  {\bibinfo  {journal} {\prb}\ }\textbf {\bibinfo {volume} {57}},\ \bibinfo
  {pages} {3433} (\bibinfo {year} {1998})}\BibitemShut {NoStop}%
\bibitem [{\citenamefont {{Karpman}}(1967)}]{Karpman1967}%
  \BibitemOpen
  \bibfield  {author} {\bibinfo {author} {\bibfnamefont {V.~I.}\ \bibnamefont
  {{Karpman}}},\ }\href@noop {} {\bibfield  {journal} {\bibinfo  {journal}
  {Pis'ma Zh.~Eksp.~Teor.~Fiz. [JETP Letters]}\ }\textbf {\bibinfo {volume}
  {6}},\ \bibinfo {pages} {277} (\bibinfo {year} {1967})}\BibitemShut {NoStop}%
\bibitem [{\citenamefont {Tai}\ \emph {et~al.}(1986)\citenamefont {Tai},
  \citenamefont {Hasegawa},\ and\ \citenamefont {Tomita}}]{Hasegawa1986}%
  \BibitemOpen
  \bibfield  {author} {\bibinfo {author} {\bibfnamefont {K.}~\bibnamefont
  {Tai}}, \bibinfo {author} {\bibfnamefont {A.}~\bibnamefont {Hasegawa}}, \
  and\ \bibinfo {author} {\bibfnamefont {A.}~\bibnamefont {Tomita}},\ }\href
  {\doibase 10.1103/PhysRevLett.56.135} {\bibfield  {journal} {\bibinfo
  {journal} {\prl}\ }\textbf {\bibinfo {volume} {56}},\ \bibinfo {pages} {135}
  (\bibinfo {year} {1986})}\BibitemShut {NoStop}%
\bibitem [{\citenamefont {{Akhmediev}}\ and\ \citenamefont
  {{Korneev}}(1986)}]{Akhmediev1986}%
  \BibitemOpen
  \bibfield  {author} {\bibinfo {author} {\bibfnamefont {N.~N.}\ \bibnamefont
  {{Akhmediev}}}\ and\ \bibinfo {author} {\bibfnamefont {V.~I.}\ \bibnamefont
  {{Korneev}}},\ }\href {\doibase 10.1007/BF01037866} {\bibfield  {journal}
  {\bibinfo  {journal} {Teoret. Mat. Fiz}\ }\textbf {\bibinfo {volume} {69}},\
  \bibinfo {pages} {1089} (\bibinfo {year} {1986})}\BibitemShut {NoStop}%
\bibitem [{\citenamefont {Matera}\ \emph {et~al.}(1993)\citenamefont {Matera},
  \citenamefont {Mecozzi}, \citenamefont {Romagnoli},\ and\ \citenamefont
  {Settembre}}]{Matera1993}%
  \BibitemOpen
  \bibfield  {author} {\bibinfo {author} {\bibfnamefont {F.}~\bibnamefont
  {Matera}}, \bibinfo {author} {\bibfnamefont {A.}~\bibnamefont {Mecozzi}},
  \bibinfo {author} {\bibfnamefont {M.}~\bibnamefont {Romagnoli}}, \ and\
  \bibinfo {author} {\bibfnamefont {M.}~\bibnamefont {Settembre}},\ }\href
  {http://www.ncbi.nlm.nih.gov/pubmed/19823425} {\bibfield  {journal} {\bibinfo
   {journal} {\ol}\ }\textbf {\bibinfo {volume} {18}},\ \bibinfo {pages} {1499}
  (\bibinfo {year} {1993})}\BibitemShut {NoStop}%
\bibitem [{\citenamefont {Kikuchi}\ \emph {et~al.}(1995)\citenamefont
  {Kikuchi}, \citenamefont {Lorattanasane}, \citenamefont {Futami},\ and\
  \citenamefont {Kaneko}}]{Kikuchi1995}%
  \BibitemOpen
  \bibfield  {author} {\bibinfo {author} {\bibfnamefont {K.}~\bibnamefont
  {Kikuchi}}, \bibinfo {author} {\bibfnamefont {C.}~\bibnamefont
  {Lorattanasane}}, \bibinfo {author} {\bibfnamefont {F.}~\bibnamefont
  {Futami}}, \ and\ \bibinfo {author} {\bibfnamefont {S.}~\bibnamefont
  {Kaneko}},\ }\href {\doibase 10.1109/68.473504} {\bibfield  {journal}
  {\bibinfo  {journal} {IEEE Photon.~Technol.~Lett.}\ }\textbf {\bibinfo
  {volume} {7}},\ \bibinfo {pages} {1378} (\bibinfo {year} {1995})}\BibitemShut
  {NoStop}%
\bibitem [{\citenamefont {Smith}\ and\ \citenamefont
  {Doran}(1996)}]{Smith1996}%
  \BibitemOpen
  \bibfield  {author} {\bibinfo {author} {\bibfnamefont {N.~J.}\ \bibnamefont
  {Smith}}\ and\ \bibinfo {author} {\bibfnamefont {N.~J.}\ \bibnamefont
  {Doran}},\ }\href {http://www.ncbi.nlm.nih.gov/pubmed/19876086} {\bibfield
  {journal} {\bibinfo  {journal} {\ol}\ }\textbf {\bibinfo {volume} {21}},\
  \bibinfo {pages} {570} (\bibinfo {year} {1996})}\BibitemShut {NoStop}%
\bibitem [{\citenamefont {Bronski}\ and\ \citenamefont {{Nathan
  Kutz}}(1996)}]{Bronski1996}%
  \BibitemOpen
  \bibfield  {author} {\bibinfo {author} {\bibfnamefont {J.~C.}\ \bibnamefont
  {Bronski}}\ and\ \bibinfo {author} {\bibfnamefont {J.}~\bibnamefont {{Nathan
  Kutz}}},\ }\href {http://www.ncbi.nlm.nih.gov/pubmed/19876210} {\bibfield
  {journal} {\bibinfo  {journal} {\ol}\ }\textbf {\bibinfo {volume} {21}},\
  \bibinfo {pages} {937} (\bibinfo {year} {1996})}\BibitemShut {NoStop}%
\bibitem [{\citenamefont {Kumar}\ \emph {et~al.}(2003)\citenamefont {Kumar},
  \citenamefont {Labruyere},\ and\ \citenamefont
  {Tchofo-Dinda}}]{TchofoDinda2003}%
  \BibitemOpen
  \bibfield  {author} {\bibinfo {author} {\bibfnamefont {A.}~\bibnamefont
  {Kumar}}, \bibinfo {author} {\bibfnamefont {A.}~\bibnamefont {Labruyere}}, \
  and\ \bibinfo {author} {\bibfnamefont {P.}~\bibnamefont {Tchofo-Dinda}},\
  }\href {\doibase 10.1016/S0030-4018(03)01290-2} {\bibfield  {journal}
  {\bibinfo  {journal} {\oc}\ }\textbf {\bibinfo {volume} {219}},\ \bibinfo
  {pages} {221} (\bibinfo {year} {2003})}\BibitemShut {NoStop}%
\bibitem [{\citenamefont {Ambomo}\ \emph {et~al.}(2008)\citenamefont {Ambomo},
  \citenamefont {Ngabireng},\ and\ \citenamefont
  {Tchofo-Dinda}}]{TchofoDinda2008}%
  \BibitemOpen
  \bibfield  {author} {\bibinfo {author} {\bibfnamefont {S.}~\bibnamefont
  {Ambomo}}, \bibinfo {author} {\bibfnamefont {C.~M.}\ \bibnamefont
  {Ngabireng}}, \ and\ \bibinfo {author} {\bibfnamefont {P.}~\bibnamefont
  {Tchofo-Dinda}},\ }\href
  {http://www.opticsinfobase.org/abstract.cfm?id=154721} {\bibfield  {journal}
  {\bibinfo  {journal} {\josab}\ }\textbf {\bibinfo {volume} {25}},\ \bibinfo
  {pages} {425} (\bibinfo {year} {2008})}\BibitemShut {NoStop}%
\bibitem [{\citenamefont {Abdullaev}\ \emph {et~al.}(1996)\citenamefont
  {Abdullaev}, \citenamefont {Darmanyan}, \citenamefont {Kobyakov},\ and\
  \citenamefont {Lederer}}]{Abdullaev1996}%
  \BibitemOpen
  \bibfield  {author} {\bibinfo {author} {\bibfnamefont {F.~K.}\ \bibnamefont
  {Abdullaev}}, \bibinfo {author} {\bibfnamefont {S.~A.}\ \bibnamefont
  {Darmanyan}}, \bibinfo {author} {\bibfnamefont {A.}~\bibnamefont {Kobyakov}},
  \ and\ \bibinfo {author} {\bibfnamefont {F.}~\bibnamefont {Lederer}},\ }\href
  {\doibase 10.1016/0375-9601(96)00504-X} {\bibfield  {journal} {\bibinfo
  {journal} {Phys. Lett. A}\ }\textbf {\bibinfo {volume} {220}},\ \bibinfo
  {pages} {213} (\bibinfo {year} {1996})}\BibitemShut {NoStop}%
\bibitem [{\citenamefont {Abdullaev}\ \emph {et~al.}(1997)\citenamefont
  {Abdullaev}, \citenamefont {Darmanyan}, \citenamefont {Bischoff},\ and\
  \citenamefont {S\o~rensen}}]{Abdullaev1997}%
  \BibitemOpen
  \bibfield  {author} {\bibinfo {author} {\bibfnamefont {F.~K.}\ \bibnamefont
  {Abdullaev}}, \bibinfo {author} {\bibfnamefont {S.~A.}\ \bibnamefont
  {Darmanyan}}, \bibinfo {author} {\bibfnamefont {S.}~\bibnamefont {Bischoff}},
  \ and\ \bibinfo {author} {\bibfnamefont {M.~P.}\ \bibnamefont {S\o~rensen}},\
  }\href {\doibase 10.1364/JOSAB.14.000027} {\bibfield  {journal} {\bibinfo
  {journal} {\josab}\ }\textbf {\bibinfo {volume} {14}},\ \bibinfo {pages} {27}
  (\bibinfo {year} {1997})}\BibitemShut {NoStop}%
\bibitem [{\citenamefont {Abdullaev}\ and\ \citenamefont
  {Garnier}(1999)}]{Abdullaev1999}%
  \BibitemOpen
  \bibfield  {author} {\bibinfo {author} {\bibfnamefont {F.~K.}\ \bibnamefont
  {Abdullaev}}\ and\ \bibinfo {author} {\bibfnamefont {J.}~\bibnamefont
  {Garnier}},\ }\href {http://www.ncbi.nlm.nih.gov/pubmed/11969851} {\bibfield
  {journal} {\bibinfo  {journal} {\pre}\ }\textbf {\bibinfo {volume} {60}},\
  \bibinfo {pages} {1042} (\bibinfo {year} {1999})}\BibitemShut {NoStop}%
\bibitem [{\citenamefont {Bauer}\ and\ \citenamefont
  {Melnikov}(1995)}]{Bauer1995}%
  \BibitemOpen
  \bibfield  {author} {\bibinfo {author} {\bibfnamefont {R.}~\bibnamefont
  {Bauer}}\ and\ \bibinfo {author} {\bibfnamefont {L.}~\bibnamefont
  {Melnikov}},\ }\href {\doibase 10.1016/0030-4018(94)00618-5} {\bibfield
  {journal} {\bibinfo  {journal} {\oc}\ }\textbf {\bibinfo {volume} {115}},\
  \bibinfo {pages} {190} (\bibinfo {year} {1995})}\BibitemShut {NoStop}%
\bibitem [{\citenamefont {Pelinovsky}\ and\ \citenamefont
  {Yang}(2004)}]{Pelinovsky2004}%
  \BibitemOpen
  \bibfield  {author} {\bibinfo {author} {\bibfnamefont {D.~E.}\ \bibnamefont
  {Pelinovsky}}\ and\ \bibinfo {author} {\bibfnamefont {J.}~\bibnamefont
  {Yang}},\ }\href {\doibase 10.1137/S0036139903422358} {\bibfield  {journal}
  {\bibinfo  {journal} {SIAM J. Appl. Math.}\ }\textbf {\bibinfo {volume}
  {64}},\ \bibinfo {pages} {1360} (\bibinfo {year} {2004})}\BibitemShut
  {NoStop}%
\bibitem [{\citenamefont {Droques}\ \emph {et~al.}(2012)\citenamefont
  {Droques}, \citenamefont {Kudlinski}, \citenamefont {Bouwmans}, \citenamefont
  {Martinelli},\ and\ \citenamefont {Mussot}}]{Droques2012}%
  \BibitemOpen
  \bibfield  {author} {\bibinfo {author} {\bibfnamefont {M.}~\bibnamefont
  {Droques}}, \bibinfo {author} {\bibfnamefont {A.}~\bibnamefont {Kudlinski}},
  \bibinfo {author} {\bibfnamefont {G.}~\bibnamefont {Bouwmans}}, \bibinfo
  {author} {\bibfnamefont {G.}~\bibnamefont {Martinelli}}, \ and\ \bibinfo
  {author} {\bibfnamefont {A.}~\bibnamefont {Mussot}},\ }\href {\doibase
  10.1364/OL.37.004832} {\bibfield  {journal} {\bibinfo  {journal} {\ol}\
  }\textbf {\bibinfo {volume} {37}},\ \bibinfo {pages} {4832} (\bibinfo {year}
  {2012})}\BibitemShut {NoStop}%
\bibitem [{\citenamefont {Droques}\ \emph {et~al.}(2013)\citenamefont
  {Droques}, \citenamefont {Kudlinski}, \citenamefont {Bouwmans}, \citenamefont
  {Martinelli},\ and\ \citenamefont {Mussot}}]{Droques2013}%
  \BibitemOpen
  \bibfield  {author} {\bibinfo {author} {\bibfnamefont {M.}~\bibnamefont
  {Droques}}, \bibinfo {author} {\bibfnamefont {A.}~\bibnamefont {Kudlinski}},
  \bibinfo {author} {\bibfnamefont {G.}~\bibnamefont {Bouwmans}}, \bibinfo
  {author} {\bibfnamefont {G.}~\bibnamefont {Martinelli}}, \ and\ \bibinfo
  {author} {\bibfnamefont {A.}~\bibnamefont {Mussot}},\ }\href {\doibase
  10.1103/PhysRevA.87.013813} {\bibfield  {journal} {\bibinfo  {journal}
  {\pra}\ }\textbf {\bibinfo {volume} {87}},\ \bibinfo {pages} {013813}
  (\bibinfo {year} {2013})}\BibitemShut {NoStop}%
\bibitem [{\citenamefont {Russell}(2003)}]{RussellScience2003}%
  \BibitemOpen
  \bibfield  {author} {\bibinfo {author} {\bibfnamefont {P.~{\relax St.J}.}\
  \bibnamefont {Russell}},\ }\href {\doibase 10.1126/science.1079280}
  {\bibfield  {journal} {\bibinfo  {journal} {Science}\ }\textbf {\bibinfo
  {volume} {299}},\ \bibinfo {pages} {358} (\bibinfo {year}
  {2003})}\BibitemShut {NoStop}%
\bibitem [{\citenamefont {Armaroli}\ and\ \citenamefont
  {Biancalana}(2012)}]{Armaroli2012}%
  \BibitemOpen
  \bibfield  {author} {\bibinfo {author} {\bibfnamefont {A.}~\bibnamefont
  {Armaroli}}\ and\ \bibinfo {author} {\bibfnamefont {F.}~\bibnamefont
  {Biancalana}},\ }\href {\doibase 10.1364/OE.20.025096} {\bibfield  {journal}
  {\bibinfo  {journal} {Opt. Express}\ }\textbf {\bibinfo {volume} {20}},\
  \bibinfo {pages} {25096} (\bibinfo {year} {2012})}\BibitemShut {NoStop}%
\bibitem [{\citenamefont {Sanders}\ \emph {et~al.}(2010)\citenamefont
  {Sanders}, \citenamefont {Verhulst},\ and\ \citenamefont
  {Murdock}}]{VerhulstBook2010}%
  \BibitemOpen
  \bibfield  {author} {\bibinfo {author} {\bibfnamefont {J.~A.}\ \bibnamefont
  {Sanders}}, \bibinfo {author} {\bibfnamefont {F.}~\bibnamefont {Verhulst}}, \
  and\ \bibinfo {author} {\bibfnamefont {J.}~\bibnamefont {Murdock}},\
  }\href@noop {} {\emph {\bibinfo {title} {Averaging Methods in Nonlinear
  Dynamical Systems (Applied Mathematical Sciences)}}}\ (\bibinfo  {publisher}
  {Springer},\ \bibinfo {year} {2010})\BibitemShut {NoStop}%
\bibitem [{\citenamefont {Agrawal}(2006)}]{AgrawalNL}%
  \BibitemOpen
  \bibfield  {author} {\bibinfo {author} {\bibfnamefont {G.}~\bibnamefont
  {Agrawal}},\ }\href
  {http://www.amazon.com/Nonlinear-Optics-Fourth-Edition-Photonics/dp/0123695163%3FSubscriptionId%3D0JYN1NVW651KCA56C102%26tag%3Dtechkie-20%26linkCode%3Dxm2%26camp%3D2025%26creative%3D165953%26creativeASIN%3D0123695163}
  {\emph {\bibinfo {title} {Nonlinear Fiber Optics, Fourth Edition (Optics and
  Photonics)}}}\ (\bibinfo  {publisher} {Academic Press},\ \bibinfo {year}
  {2006})\BibitemShut {NoStop}%
\bibitem [{\citenamefont {Rothenberg}(1990)}]{Rothenberg1990a}%
  \BibitemOpen
  \bibfield  {author} {\bibinfo {author} {\bibfnamefont {J.~E.}\ \bibnamefont
  {Rothenberg}},\ }\href {http://link.aps.org/doi/10.1103/PhysRevA.42.682}
  {\bibfield  {journal} {\bibinfo  {journal} {\pra}\ }\textbf {\bibinfo
  {volume} {42}},\ \bibinfo {pages} {682} (\bibinfo {year} {1990})}\BibitemShut
  {NoStop}%
\bibitem [{\citenamefont {Drummond}\ \emph {et~al.}(1990)\citenamefont
  {Drummond}, \citenamefont {Kennedy}, \citenamefont {Dudley}, \citenamefont
  {Leonhardt},\ and\ \citenamefont {Harvey}}]{Drummond1990}%
  \BibitemOpen
  \bibfield  {author} {\bibinfo {author} {\bibfnamefont {P.}~\bibnamefont
  {Drummond}}, \bibinfo {author} {\bibfnamefont {T.}~\bibnamefont {Kennedy}},
  \bibinfo {author} {\bibfnamefont {J.}~\bibnamefont {Dudley}}, \bibinfo
  {author} {\bibfnamefont {R.}~\bibnamefont {Leonhardt}}, \ and\ \bibinfo
  {author} {\bibfnamefont {J.}~\bibnamefont {Harvey}},\ }\href {\doibase
  10.1016/0030-4018(90)90110-F} {\bibfield  {journal} {\bibinfo  {journal}
  {Opt.~Commun.}\ }\textbf {\bibinfo {volume} {78}},\ \bibinfo {pages} {137 }
  (\bibinfo {year} {1990})}\BibitemShut {NoStop}%
\bibitem [{\citenamefont {Rothenberg}(1991)}]{Rothenberg1991}%
  \BibitemOpen
  \bibfield  {author} {\bibinfo {author} {\bibfnamefont {J.~E.}\ \bibnamefont
  {Rothenberg}},\ }\href {http://www.ncbi.nlm.nih.gov/pubmed/19773823}
  {\bibfield  {journal} {\bibinfo  {journal} {\ol}\ }\textbf {\bibinfo {volume}
  {16}},\ \bibinfo {pages} {18} (\bibinfo {year} {1991})}\BibitemShut {NoStop}%
\bibitem [{\citenamefont {Wabnitz}(1988)}]{Wabnitz1988}%
  \BibitemOpen
  \bibfield  {author} {\bibinfo {author} {\bibfnamefont {S.}~\bibnamefont
  {Wabnitz}},\ }\href {http://adsabs.harvard.edu/abs/1988PhRvA..38.2018W}
  {\bibfield  {journal} {\bibinfo  {journal} {\pra}\ }\textbf {\bibinfo
  {volume} {38}},\ \bibinfo {pages} {2018} (\bibinfo {year}
  {1988})}\BibitemShut {NoStop}%
\bibitem [{\citenamefont {Murdoch}\ \emph {et~al.}(1997)\citenamefont
  {Murdoch}, \citenamefont {Leonhardt}, \citenamefont {Harvey},\ and\
  \citenamefont {Kennedy}}]{Murdoch1997}%
  \BibitemOpen
  \bibfield  {author} {\bibinfo {author} {\bibfnamefont {S.~G.}\ \bibnamefont
  {Murdoch}}, \bibinfo {author} {\bibfnamefont {R.}~\bibnamefont {Leonhardt}},
  \bibinfo {author} {\bibfnamefont {J.~D.}\ \bibnamefont {Harvey}}, \ and\
  \bibinfo {author} {\bibfnamefont {T.~A.~B.}\ \bibnamefont {Kennedy}},\ }\href
  {\doibase 10.1364/JOSAB.14.001816} {\bibfield  {journal} {\bibinfo  {journal}
  {\josab}\ }\textbf {\bibinfo {volume} {14}},\ \bibinfo {pages} {1816}
  (\bibinfo {year} {1997})}\BibitemShut {NoStop}%
\bibitem [{\citenamefont {Agrawal}\ \emph {et~al.}(1989)\citenamefont
  {Agrawal}, \citenamefont {Baldeck},\ and\ \citenamefont
  {Alfano}}]{Agrawal1989}%
  \BibitemOpen
  \bibfield  {author} {\bibinfo {author} {\bibfnamefont {G.~P.}\ \bibnamefont
  {Agrawal}}, \bibinfo {author} {\bibfnamefont {P.~L.}\ \bibnamefont
  {Baldeck}}, \ and\ \bibinfo {author} {\bibfnamefont {R.~R.}\ \bibnamefont
  {Alfano}},\ }\href {\doibase 10.1103/PhysRevA.39.3406} {\bibfield  {journal}
  {\bibinfo  {journal} {\pra}\ }\textbf {\bibinfo {volume} {39}},\ \bibinfo
  {pages} {3406} (\bibinfo {year} {1989})}\BibitemShut {NoStop}%
\bibitem [{\citenamefont {Millot}\ \emph {et~al.}(2002)\citenamefont {Millot},
  \citenamefont {Pitois},\ and\ \citenamefont {Tchofo-Dinda}}]{Millot2002}%
  \BibitemOpen
  \bibfield  {author} {\bibinfo {author} {\bibfnamefont {G.}~\bibnamefont
  {Millot}}, \bibinfo {author} {\bibfnamefont {S.}~\bibnamefont {Pitois}}, \
  and\ \bibinfo {author} {\bibfnamefont {P.}~\bibnamefont {Tchofo-Dinda}},\
  }\href {\doibase 10.1364/JOSAB.19.000454} {\bibfield  {journal} {\bibinfo
  {journal} {\josab}\ }\textbf {\bibinfo {volume} {19}},\ \bibinfo {pages}
  {454} (\bibinfo {year} {2002})}\BibitemShut {NoStop}%
\bibitem [{Note1()}]{Note1}%
  \BibitemOpen
  \bibinfo {note} {If $\protect \Tr {H_0}\not =0$ we could make a change of
  variables to transform it into a traceless matrix}\BibitemShut {NoStop}%
\bibitem [{\citenamefont {Biancalana}\ and\ \citenamefont
  {Skryabin}(2004)}]{Biancalana2004e}%
  \BibitemOpen
  \bibfield  {author} {\bibinfo {author} {\bibfnamefont {F.}~\bibnamefont
  {Biancalana}}\ and\ \bibinfo {author} {\bibfnamefont {D.~V.}\ \bibnamefont
  {Skryabin}},\ }\href {\doibase 10.1088/1464-4258/6/4/002} {\bibfield
  {journal} {\bibinfo  {journal} {J. Opt. A--Pure Appl.~Opt.}\ }\textbf
  {\bibinfo {volume} {6}},\ \bibinfo {pages} {301} (\bibinfo {year}
  {2004})}\BibitemShut {NoStop}%
\bibitem [{\citenamefont {Kibler}\ \emph {et~al.}(2004)\citenamefont {Kibler},
  \citenamefont {Billet}, \citenamefont {Dudley}, \citenamefont {Windeler},\
  and\ \citenamefont {Millot}}]{Kibler2004}%
  \BibitemOpen
  \bibfield  {author} {\bibinfo {author} {\bibfnamefont {B.}~\bibnamefont
  {Kibler}}, \bibinfo {author} {\bibfnamefont {C.}~\bibnamefont {Billet}},
  \bibinfo {author} {\bibfnamefont {J.~M.}\ \bibnamefont {Dudley}}, \bibinfo
  {author} {\bibfnamefont {R.~S.}\ \bibnamefont {Windeler}}, \ and\ \bibinfo
  {author} {\bibfnamefont {G.}~\bibnamefont {Millot}},\ }\href
  {http://www.ncbi.nlm.nih.gov/pubmed/15357354} {\bibfield  {journal} {\bibinfo
   {journal} {\ol}\ }\textbf {\bibinfo {volume} {29}},\ \bibinfo {pages} {1903}
  (\bibinfo {year} {2004})}\BibitemShut {NoStop}%
\bibitem [{\citenamefont {Wong}\ \emph {et~al.}(2005)\citenamefont {Wong},
  \citenamefont {Chen}, \citenamefont {Murdoch}, \citenamefont {Leonhardt},
  \citenamefont {Harvey}, \citenamefont {Joly}, \citenamefont {Knight},
  \citenamefont {Wadsworth},\ and\ \citenamefont {Russell}}]{Wong2005}%
  \BibitemOpen
  \bibfield  {author} {\bibinfo {author} {\bibfnamefont {G.~K.~L.}\
  \bibnamefont {Wong}}, \bibinfo {author} {\bibfnamefont {A.~Y.~H.}\
  \bibnamefont {Chen}}, \bibinfo {author} {\bibfnamefont {S.~G.}\ \bibnamefont
  {Murdoch}}, \bibinfo {author} {\bibfnamefont {R.}~\bibnamefont {Leonhardt}},
  \bibinfo {author} {\bibfnamefont {J.~D.}\ \bibnamefont {Harvey}}, \bibinfo
  {author} {\bibfnamefont {N.~Y.}\ \bibnamefont {Joly}}, \bibinfo {author}
  {\bibfnamefont {J.~C.}\ \bibnamefont {Knight}}, \bibinfo {author}
  {\bibfnamefont {W.~J.}\ \bibnamefont {Wadsworth}}, \ and\ \bibinfo {author}
  {\bibfnamefont {P.~{\relax St.J}.}\ \bibnamefont {Russell}},\ }\href
  {\doibase 10.1364/JOSAB.22.002505} {\bibfield  {journal} {\bibinfo  {journal}
  {\josab}\ }\textbf {\bibinfo {volume} {22}},\ \bibinfo {pages} {2505}
  (\bibinfo {year} {2005})}\BibitemShut {NoStop}%
\bibitem [{\citenamefont {Chen}\ \emph {et~al.}(2006)\citenamefont {Chen},
  \citenamefont {Wong}, \citenamefont {Murdoch}, \citenamefont {Kruhlak},
  \citenamefont {Leonhardt}, \citenamefont {Harvey}, \citenamefont {Joly},\
  and\ \citenamefont {Knight}}]{Chen2006}%
  \BibitemOpen
  \bibfield  {author} {\bibinfo {author} {\bibfnamefont {J.~S.}\ \bibnamefont
  {Chen}}, \bibinfo {author} {\bibfnamefont {G.~K.}\ \bibnamefont {Wong}},
  \bibinfo {author} {\bibfnamefont {S.~G.}\ \bibnamefont {Murdoch}}, \bibinfo
  {author} {\bibfnamefont {R.~J.}\ \bibnamefont {Kruhlak}}, \bibinfo {author}
  {\bibfnamefont {R.}~\bibnamefont {Leonhardt}}, \bibinfo {author}
  {\bibfnamefont {J.~D.}\ \bibnamefont {Harvey}}, \bibinfo {author}
  {\bibfnamefont {N.~Y.}\ \bibnamefont {Joly}}, \ and\ \bibinfo {author}
  {\bibfnamefont {J.~C.}\ \bibnamefont {Knight}},\ }\href {\doibase
  10.1364/OL.31.000873} {\bibfield  {journal} {\bibinfo  {journal} {\ol}\
  }\textbf {\bibinfo {volume} {31}},\ \bibinfo {pages} {873} (\bibinfo {year}
  {2006})}\BibitemShut {NoStop}%
\bibitem [{\citenamefont {Kudlinski}\ \emph {et~al.}(2013)\citenamefont
  {Kudlinski}, \citenamefont {Bendahmane}, \citenamefont {Labat}, \citenamefont
  {Virally}, \citenamefont {Murray}, \citenamefont {Kelleher},\ and\
  \citenamefont {Mussot}}]{Kudlinski2013}%
  \BibitemOpen
  \bibfield  {author} {\bibinfo {author} {\bibfnamefont {A.}~\bibnamefont
  {Kudlinski}}, \bibinfo {author} {\bibfnamefont {A.}~\bibnamefont
  {Bendahmane}}, \bibinfo {author} {\bibfnamefont {D.}~\bibnamefont {Labat}},
  \bibinfo {author} {\bibfnamefont {S.}~\bibnamefont {Virally}}, \bibinfo
  {author} {\bibfnamefont {R.~T.}\ \bibnamefont {Murray}}, \bibinfo {author}
  {\bibfnamefont {E.~J.~R.}\ \bibnamefont {Kelleher}}, \ and\ \bibinfo {author}
  {\bibfnamefont {A.}~\bibnamefont {Mussot}},\ }\href {\doibase
  10.1364/OE.21.008437} {\bibfield  {journal} {\bibinfo  {journal} {Opt.
  Express}\ }\textbf {\bibinfo {volume} {21}},\ \bibinfo {pages} {8437}
  (\bibinfo {year} {2013})}\BibitemShut {NoStop}%
\end{thebibliography}%

\end{document}